\documentclass[12pt,preprint]{aastex}
\slugcomment{To appear in the Astrophysical Journal}

\lefthead{Skinner et al.}
\righthead{RY Tau}

\begin{document}

\title{Chandra and XMM-Newton X-Ray Observations of the Hyperactive T Tauri Star  RY Tau}

\author{Stephen L. Skinner\footnote{Center for Astrophysics and 
Space Astronomy (CASA), Univ. of Colorado,
Boulder, CO, USA 80309-0389; stephen.skinner@colorado.edu},
Marc Audard\footnote{Dept. of Astronomy, 
University of Geneva, Ch. d'Ecogia 16, CH-1290 Versoix, Switzerland; marc.audard@unige.ch}, and
Manuel  G\"{u}del\footnote{Dept. of Astrophysics, Univ. of Vienna, 
T\"{u}rkenschanzstr. 17,  A-1180 Vienna, Austria; 
manuel.guedel@univie.ac.at}}

%
\newcommand{\ltsimeq}{\raisebox{-0.6ex}{$\,\stackrel{\raisebox{-.2ex}%
{$\textstyle<$}}{\sim}\,$}}
%
\newcommand{\gtsimeq}{\raisebox{-0.6ex}{$\,\stackrel{\raisebox{-.2ex}%
{$\textstyle>$}}{\sim}\,$}}

\begin{abstract}
\small{
We present results of pointed X-ray observations of the accreting
jet-driving  T Tauri star RY Tau using {\em Chandra} and {\em XMM-Newton}. 
We obtained  high-resolution grating spectra
and excellent-quality CCD spectra and light curves with the objective
of identifying the physical mechanisms underlying RY Tau's
bright X-ray emission. Grating spectra reveal numerous 
emission lines spanning a broad range of temperature 
superimposed on a hot continuum. The X-ray emission
measure distribution is dominated by very hot plasma at
T$_{hot}$ $\sim$ 50 MK but higher temperatures were 
present during flares. A weaker cool plasma component
is also present as revealed by low-temperature lines such
as O VIII.  X-ray light curves show complex variability
consisting of short-duration ($\sim$hours) superhot flares accompanied
by fluorescent Fe emission at 6.4 keV superimposed on a slowly-varying 
($\sim$one day) component that may be tied to stellar rotation. The hot flaring
component is undoubtedly of magnetic (e.g. coronal) origin.
Soft and hard-band light curves undergo similar slow 
variability implying that at least some of the cool plasma shares a 
common magnetic origin with the hot plasma. Any contribution to the
X-ray emission from cool shocked plasma is small compared to the 
dominant hot component but production of individual low-temperature
lines such as O VIII in an accretion shock is not ruled out.
}
\end{abstract}





\section{Introduction}
Classical T Tauri stars (cTTS) are  low-mass pre-main 
sequence stars that are  accreting from disks and
continue to be high-interest objects for their remarkable
range of activity (Bertout 1989) and as potential host stars 
for nascent protoplanetary systems.
They are intense X-ray emitters whose X-ray luminosities
typically exceed that of the contemporary Sun
by $\sim$2-3 orders of magnitude (Feigelson \& Montmerle 1999).
The intense X-ray emission heats and
ionizes the inner disk, altering disk chemistry and influencing 
the physical environment in which planet formation 
occurs (Glassgold et al. 1997; Shang et al. 2002).

Early  studies of cTTS suggested  that their
X-ray emission is dominated by magnetic (e.g. coronal)
processes analogous to scaled-up solar activity
(Feigelson \& Montmerle 1999). This emission is often variable
and can display huge magnetic reconnection flares
characterized by rapid changes ($\sim$hours to days)
in the observed X-ray flux and  elevated plasma temperatures 
reaching  extreme values  T $\gtsimeq$ 100 MK. 

However, more recent observational work has shown that 
other physical processes  also contribute to the X-ray
emission of cTTS. Cool plasma (T $<$ 3 MK) that 
is thought to arise in the post-shock zone of an accretion shock 
at or near the stellar surface has been detected in a
few cTTS, most notably TW Hya (Kastner et al. 2002; 
Stelzer \& Schmitt 2004; Brickhouse et al. 2010). There 
is also accumulating evidence that faint soft X-rays are produced 
in collimated high-speed jets within a few  arcseconds of 
the central star.  Such cTTS X-ray jet sources 
include the well-studied object DG Tau (G\"{u}del et al. 2005; 2008) 
and possible X-ray jet detections in RY Tau (Skinner, Audard, 
\& G\"{u}del 2011; hereafter SAG11) and RW Aur (Skinner \& G\"{u}del 2014).
X-ray jet emission was also observed following an optical 
outburst in the unusual pre-main sequence binary Z CMa 
(Stelzer et al. 2009). Soft X-ray emission that likely
originates in shock-heated material has been reported 
in some Herbig-Haro objects of which a few examples are
HH 2 (Pravdo et al. 2001), HH 80/81 (Pravdo et al. 2004),
HH 154 (Favata et al. 2002, 2006; Bally et al. 2003),
and HH 210 (Grosso et al. 2006).

X-ray jet emission is usually recognized in high-resolution
images as $\sim$arcsecond-scale structure extending outward  
from the unresolved central star along one or (rarely) both of
the optical jet axes.
{\em Chandra}'s excellent spatial resolution has proven to 
be especially useful in identifying such faint extended  
jet structure. On the other hand, coronal and accretion shock 
X-ray emission originate near the stellar surface and cannot 
be spatially resolved with existing X-ray telescopes. But 
analysis of individual emission lines in high-resolution X-ray 
grating spectra can potentially distinguish  between cool dense 
plasma in accretion shocks (T $\sim$ few MK; 
n$_{e}$ $\gtsimeq$ 10$^{12}$ cm$^{-3}$) and hotter lower density coronal plasma
(typically  n$_{e}$ $\ltsimeq$ 10$^{12}$ cm$^{-3}$), as
clearly demonstrated in the case of TW Hya.

Relatively few cTTS are bright enough in X-rays to permit 
good-quality grating spectra to be acquired in reasonable
exposure  times. We identified the accreting jet-driving cTTS
RY Tau as a promising candidate for grating spectroscopy on the 
basis of bright  X-ray emission detected in a 2009 {\em Chandra} ACIS-S 
observation (SAG11). Moderate resolution CCD 
spectra revealed multi-temperature plasma consisting of a superhot
flaring component (typical of coronal emission) and 
cooler plasma of uncertain origin. Deconvolved soft-band images 
showed some evidence of extension outward along the 
blueshifted optical jet.

We present here new X-ray  observations of
the cTTS RY Tau obtained with {\em Chandra} and
{\em XMM-Newton}. These observations include the first
high-resolution  grating spectra of RY Tau as well as
high-quality undispersed CCD spectra and X-ray light curves.
Our main objective  is to obtain 
a more complete picture  of the properties of the 
X-ray emitting plasma (i.e. temperature, electron density, 
emission measure distribution) in order to differentiate
between hot magnetically-confined plasma (e.g. coronal emission)
and cooler plasma that may originate in  shocks.

\begin{deluxetable}{lcccccccc}
\tabletypesize{\scriptsize}
\tablewidth{0pt}
\tablecaption{Properties of RY Tau}
\tablehead{
           \colhead{Sp. Type}               &
           \colhead{M$_{*}$}            &
           \colhead{V}            &
           \colhead{A$_{\rm V}$}            &
           \colhead{L$_{*}$}            &
           \colhead{L$_{bol}$}            &
           \colhead{log L$_{x}$}     &
           \colhead{log $\dot{M}_{acc}$}            &
           \colhead{d}             \\
           \colhead{}                   &
           \colhead{(M$_{\odot}$)}         &
           \colhead{(mag)}                   &
           \colhead{(mag)}                   &
           \colhead{(L$_{\odot}$)}           &
           \colhead{(L$_{\odot}$)}           &
           \colhead{(ergs s$^{-1}$)}         &
           \colhead{(M$_{\odot}$ yr$^{-1}$)}     &
           \colhead{(pc)}              
}
\startdata
 F8 III - G3 IV  & 1.7 - 2.0   &  9.3 - 11 (v) & 2.2$\pm$0.2    & 7.0 - 8.8  & 15.3 & 30.5 - 31.2(v)   & $-$7.3$\pm$0.3     & 134     \\
\enddata
\tablecomments{Data are from Agra-Amboage et al. (2009), Calvet et al. (2004), Holtzman et al. (1986), Kenyon \& Hartmann (1995),
and Schegerer et al. (2008). The stellar luminosity L$_{*}$ and bolometric luminosity  L$_{bol}$ are
based on the values L$_{*}$ = 7.6 - 9.6 L$_{\odot}$ and L$_{bol}$ = 16.7 L$_{\odot}$ at d = 140 pc
(Calvet et al. 2004, Kenyon \& Hartmann 1995) and have been normalized to d = 134 pc.
L$_{x}$ range is based on this work.
 Distance is from Bertout et al. (1999). Variable parameters are denoted by (v). 
 Spectral type is uncertain (Holtzman, Herbst, \& Booth 1986).}
\end{deluxetable}

\section{RY Tau}

RY Tau is an optically variable  cTTS 
lying in the Taurus dark cloud (Bertout, Robichon, \& Arenou 1999). 
Its properties are summarized in Table 1. Its mass 
M$_{*}$ $\approx$ 2 M$_{\odot}$ is relatively
high for a TTS and it rotates rapidly at 
$v$sin$i$  = 52 $\pm$ 2 km s$^{-1}$ (Petrov et al. 1999). 
It is accreting from a disk (Schegerer et al. 2008; Angra-Amboage
et al. 2009) and mass-loss is clearly 
evident in the form of a bipolar jet known as HH 938
(St.-Onge \& Bastien 2008) and a wind
(G\'{o}mez de Castro \& Verdugo 2007). Optical images have
traced the jet and associated H$\alpha$ knots outward to 
a separation of 31$''$ from the star along
P.A. $\approx$ 295$^{\circ}$ (measured east from north), and
the fainter counterjet is  visible out to 3.$'$5 from the star. 
Binarity is suspected based on {\em Hipparcos} photocenter
motions (Bertout et al. 1999) but no companion has yet been 
found (Leinert et al. 1993; Schegerer et al. 2008; Pott et al. 2010).
Searches for periodic optical variability have reported periods
of $\sim$years but no stable short period of $\sim$days that might be associated
with stellar rotation has yet been confirmed (Bouvier et al. 1993; 
Ismailov \& Adygezalzade 2012 and references therein; Zajtseva 2010).

\section{Summary of Previous XMM-Newton and Chandra Observations}

{\em 2000}:~RY Tau was  captured 8.5$'$ off-axis in a September 2000
{\em XMM-Newton} observation targeted on HDE 283572 (51 ks; ObsId 0101440701).
This archived observation was included in the 
{\em XMM-Newton Extended Survey of the Taurus Molecular Cloud} (XEST) 
as reported by G\"{u}del et al. (2007a). RY Tau was cataloged as
XEST source 21-038. Fits of its CCD spectra with an
absorbed 2T thermal plasma model gave N$_{\rm H}$ = 7.6 ($+$0.9,$-$1.0) 
$\times$ 10$^{21}$ cm$^{-2}$,
kT$_{cool}$ = 0.5 keV, kT$_{hot}$ = 3.2 keV, and
log L$_{x}$(0.3 - 10 keV)  = 30.72 ergs s$^{-1}$ (d = 140 pc).
No large-amplitude variability was reported.

{\em 2003}:~RY Tau was captured off-axis in a October 2003 {\em Chandra} 
ACIS-S/HETG grating observation targeted on HDE 283572 (ObsId 3756). 
The zero-order ACIS-S spectrum and light curve of RY Tau were analyzed by
Audard et al. (2005) but the grating spectrum  was not analyzed
due to lack of reliable calibration for far off-axis sources.
The $\approx$100 ks X-ray light curve  
showed a slow decline punctuated by
a short-duration low-amplitude flare.  The ACIS-S
spectrum was fitted with an absorbed two-temperature (2T) thermal plasma
model with temperatures kT$_{cool}$ = 1.0 keV, kT$_{hot}$ = 3.9 keV,
absorption N$_{\rm H}$ = 5 ($\pm$1) $\times$ 10$^{21}$ cm$^{-2}$, 
subsolar metallicity $Z$ = 0.3 Z$_{\odot}$, and X-ray luminosity
log L$_{x}$ (0.1 - 10 keV) = 30.88 ergs s$^{-1}$ at d = 140 pc. 
Fluorescent Fe emission was detected at $\approx$6.4 keV.

{\em 2009}:~We obtained a 56 ks  observation with RY Tau on-axis
using {\em Chandra} ACIS-S without gratings in
December 2009 (ObsId 10991). The
star was  detected as a bright variable X-ray source (SAG11).
An absorbed 2T thermal plasma model gave
N$_{\rm H}$ = 5.5 ($+$1.3,$-$0.8) $\times$ 10$^{21}$ cm$^{-2}$,
kT$_{cool}$ = 0.6 keV, kT$_{hot}$ = 4.8 keV, $Z$ = 0.25 Z$_{\odot}$,
and log L$_{x}$(0.2 - 10 keV)  = 30.67 ergs s$^{-1}$ (d = 134 pc).
For comparison with the above studies, the latter value equates
to log L$_{x}$ = 30.71 ergs s$^{-1}$ at d = 140 pc.
Deconvolved soft-band (0.2 - 2 keV) ACIS-S images 
revealed faint extended structure overlapping the inner blushifted optical
jet and traced out to a separation of $\approx$1.$''$7 from the
star. A known asymmetry in the {\em Chandra} point-spread function
may contribute to some of the structure inside one arcsecond.
Five other  X-ray sources were detected within one arcminute of
RY Tau but all were quite faint.

\begin{deluxetable}{llll}
\tabletypesize{\scriptsize}
\tablewidth{0pc}
\tablecaption{RY Tau X-ray Observations }
\tablehead{
\colhead{Telescope} &
\colhead{XMM-Newton} &
\colhead{Chandra} &
\colhead{Chandra} \\
}
\startdata
Date                              & 2013 Aug. 21-22  & 2014 Oct. 22-23   & 2014 Oct. 23-24   \nl
ObsId                             & 0722320101       & 15722             & 17539          \nl
Start Time (UTC)                  & 04:34            & 02:06             & 21:37          \nl
Stop  Time (UTC)                  & 11:13            & 03:06             & 04:49         \nl
Duration (ks)                    & 92.25\tablenotemark{a}            & 85.5\tablenotemark{b}              & 23.9\tablenotemark{b}         \nl
Prime Instrument                  & RGS              & ACIS-S/HETG       & ACIS-S/HETG       \nl
\enddata
\tablenotetext{a}{Usable exposure time (per RGS) after removing high background interval.
 The usable exposure times for EPIC were 90 ks (pn) and 100 ks (per MOS).}
\tablenotetext{b}{Livetime. Data obtained in faint timed event mode with 1.7 s frame time.}
\end{deluxetable}

\clearpage

\section{Observations and Data Reduction}

The X-ray observations are summarized in Table 2.

\subsection{XMM-Newton}

The {\em XMM-Newton} data were obtained in a single 
observation spanning $\approx$31 hours in August 2013.  
The primary instrument was
the Reflection Grating Spectrometer (RGS) which consists
of two spectrometers (RGS1 and RGS2) with similar spectral 
coverage over the $\approx$0.35 - 2.5 keV energy range
but with some gaps\footnote{RGS1 lacks coverage in the range
10.6 - 13.8 \AA~(0.90 - 1.17 keV) which includes the Ne IX He-like triplet.
RGS2 lacks coverage in the range 20.0 - 24.1 \AA~(0.51 - 0.62 keV) which includes 
the O VII He-like triplet. For further details on RGS see
~http://xmm.esac.esa.int/external/xmm\_user\_support/documentation/uhb/rgs.html .}.
The RGS effective area in first order reaches a maximum of $\approx$50 - 60 cm$^{2}$
per RGS at $\approx$15 \AA ~($\approx$0.8 keV), thus providing better sensitivity
at lower energies $\ltsimeq$1 keV than the  {\em Chandra} High Energy Transmission 
Grating (HETG).  RGS first order spectral resolution at 12 \AA ~($\approx$1 keV) is
$\Delta\lambda$ $\approx$ 0.055 - 0.07 \AA ~(FWHM). In  addition,
{\em XMM-Newton}  provided  CCD imaging spectroscopy 
from the European Photon Imaging Camera (EPIC) which was
used with the Medium optical filter.
The EPIC consists of the pn camera and two nearly-identical
MOS cameras (MOS1 and MOS2; Turner et al. 2001). The EPIC cameras 
provide energy coverage in the range E $\approx$ 0.2 - 15 keV with 
energy resolution (FWHM) at 1 keV of $\Delta E$ $\approx$ 70 eV (MOS)
and $\Delta E$ $\approx$ 100 eV (pn).

Data were reduced using the {\em XMM-Newton} Science Analysis System
(SAS vers. 13.5.0). Event files were time-filtered to remove
data acquired during a background flare at the end of the observation,
resulting in $\approx$90 - 100 ks of usable exposure per instrument (Table 2).
EPIC spectra and energy-filtered light curves were extracted from a circular
region of radius 20$''$ centered on the source and background was
extracted from source-free regions near RY Tau. 
The SAS task $rgsproc$ produced edited RGS event files
that excluded the high-background time interval.
RGS response matrix files (RMFs) were 
generated using the SAS task $rgsrmfgen$. Spectra were analyzed
with Sherpa in CIAO version 4.6\footnote{Further information on
Chandra Interactive Analysis of Observations (CIAO) software can be found at
http://asc.harvard.edu/ciao .}
and with XSPEC version 12.8.2. Timing analysis was performed with
XRONOS v. 5.22.  XSPEC and XRONOS  are included in the HEASOFT
XANADU\footnote{http://heasarc.nasa.gov/lheasoft/xanadu/ } software package.

\subsection{Chandra}

The {\em Chandra} data were acquired with the HETG and ACIS-S detector
array in October 2014 in two observations spanning $\approx$24 hours and 
$\approx$7 hours with an intervening  gap of $\approx$18.5 hours (Table 2).
The combined livetime of the two observations was 109.4 ks, of which 85.5 ks was
obtained in the first observation. The  nominal pointing
direction and roll angle were nearly  identical for the two observations.
Five of the six CCDs in the ACIS-S array were enabled (chip S0 was turned off).

We analyzed dispersed first-order  HETG spectra from the 
Medium Energy Grating (MEG) and High Energy Grating (HEG)
as well as undispersed (zero-order) ACIS-S spectra and X-ray light 
curves. The data were processed using
standard science threads in CIAO version 4.6 and associated calibration
data from  CALDB version 4.6.1.1. Our analysis is based
on Level 2 event files  provided by {\em Chandra} X-ray Center
(CXC) standard processing. 

The HETG\footnote{Further information on {\em Chandra}
instrumentation can be found at http://cxc.harvard.edu/proposer/POG/ .}
provides spectral coverage over the $\approx$0.4 - 10 keV
($\approx$1.3  - 30 \AA~) range. Background rejection is
high in HETG because of event order discrimination by ACIS-S,
in sharp contrast to the much higher background in {\em XMM-Newton}
RGS spectra.  In addition, HETG provides spectral coverage above 2.5 keV where RGS
coverage is lacking and is thus crucial for diagnosing hot plasma.
Spectral resolution in first order   is
$\Delta\lambda$ $\approx$ 0.023 \AA~ for MEG and $\Delta\lambda$ $\approx$ 0.012 \AA~ for HEG.
The separate $+$1 and $-1$ order MEG spectra were co-added for analysis,
and similarly for HEG. The effective area of MEG is maximum between
1 - 2 keV and HEG is maximum between 2 - 3 keV. The effective 
area of both gratings is lower than RGS at energies $<$1 keV.
The HEG has superior spectral resolution but the
MEG provides more total counts and higher  signal-to-noise ratio  
in first-order spectra for this observation. Our {\em Chandra}
grating analysis thus emphasizes the MEG spectrum for the 
longer first exposure (ObsId 15722)  when most of the source counts 
were accumulated.

The zero-order  ACIS-S CCD spectra and X-ray
light curves  were extracted from a circular region of radius 2$''$
centered on the RY Tau source peak. Background is negligible.
ACIS-S provides energy coverage from $\approx$0.4 - 10 keV
and its intrinsic energy resolution
is $\Delta E$ $\approx$ 120 eV (FWHM) at E = 1 keV.
Unlike our previous 2009 {\em Chandra} ACIS-S observation (without gratings), 
no 0-order image deconvolution was performed for the new ACIS-S/HETG 
observation discussed here. The point-spread-function (PSF) needed 
to produce the deconvolved image is spectral-dependent and  
the spectrum varied considerably throughout the 
observation (Sec. 5.2). In addition, the gratings disperse a significant
fraction of the incoming photons, especially at energies below 2 keV,
thereby reducing the sensitivity in 0-order to any faint soft ($<$2 keV)
extended X-ray jet emission at small offsets from the star.

\section{Results}

\subsection{X-ray Light Curves and Variability }
The {\em XMM-Newton} EPIC  broad-band light curves (Fig. 1-top)
are clearly variable during the first half of
the observation. Two flare-like outbursts occurred
within the time interval $\sim$20 - 45 ks after the start
of the observation. EPIC spectra
extracted during these flare intervals (Sec. 5.2) give 
substantially higher mean plasma temperatures than obtained 
from spectra extracted after removing the flares.
Energy-filtered pn light curves
(Fig. 1-bottom) show no significant variability in the 
very soft band (0.3 - 1 keV) but have a high probability of 
variability P$_{var}$ $>$ 0.99 in the medium (1 - 2 keV)
and hard (2 - 8 keV) bands. The medium-band light curve
reveals a slow increase in count rate during the 
observation with an average rate of 82$\pm$2 c ks$^{-1}$
in the first half and 92$\pm$2 c ks$^{-1}$ in the 
second half. Flare-like variability is most obvious
in the  hard band.

The {\em Chandra} ACIS-S 0-order  light 
curves are also variable.  As shown in Fig. 2-top left, a slow 
decrease in count rate occurred during the first 40 ks 
of the first observation (ObsId 15722), followed by a moderate slow increase. 
The slow decay and subsequent rise are clearly detected in the very soft,
medium,  and hard energy bands (Fig. 2-bottom left). An exponential  fit of 
the decay portion of the broad-band (0.3 - 8 keV) ACIS-S 
light curve gives an e-folding time
of 27.5 ks. The plasma temperature and X-ray flux decreased during
the decay phase and then began to increase again after the 
light curve minimum, as discussed further in Sec. 6. 

The average count rate and plasma temperature were lower in
the second {\em Chandra} observation (ObsId 17539). A slow  
decrease  in the broad-band (0.3 - 8 keV) and hard-band
(2 - 8 keV) count rates is clearly visible (Fig. 2-right).
There are insufficient counts to construct a separate 
very-soft band (0.3 - 1 keV) light curve so a 0.3 - 2 keV
band light curve was generated to search for low-energy
variability. This light curve is nearly constant during 
the first half of the observation but declines during the
second half (Fig. 2-bottom right). The respective count
rates in the first and second halves were 9.42 $\pm$ 0.89 c ks$^{-1}$
and 8.08 $\pm$ 1.86 c ks$^{-1}$ (1$\sigma$ uncertainties).
A $\chi^2$ test applied to the second half of the observation
gives a variability probability  P$_{var}$ = 0.60 which 
is suggestive of  variability but not conclusive.

\subsection{Undispersed CCD X-ray Spectra}

Figure 3 shows the {\em XMM-Newton}  EPIC pn spectra, which 
provide higher count rates and better sensitivity at low energies 
than MOS spectra.
Spectra for the flare and non-flare time intervals are overlaid.
The flare spectrum extracted during the 25 ks flare interval
(as marked in the bottom panel of Fig. 1) is clearly harder
as evidenced by elevated continuum emission
above $\sim$2 keV. The ratio H/S of hard-band (2 - 8 keV) 
to soft-band (0.3 - 2 keV) EPIC pn counts during the flare
segment was H/S = 1.10 as compared to H/S = 0.67 during
the non-flare time interval. 
The Fe K line complex near 6.7 keV (Fe XXV) which
originates in very hot plasma (maximum emissivity temperature
T$_{e,max}$ $\approx$ 63 MK) is present in both the flare 
and flare-excluded segments.
Interestingly, an emission feature at $\approx$6.4 keV
is present in the flare spectrum (Fig. 3-bottom).
A Gaussian fit of this feature gives a centroid energy 
E = 6.395 $\pm$ 0.005 (1$\sigma$) keV. This line is
fluorescent emission from neutral or near-neutral Fe irradiated by
the hard flaring X-ray source.

Figure 4 shows the undispersed ACIS-S spectra extracted during
the first {\em Chandra} observation (ObsId 15722).
The spectrum corresponding to the first 35 ks of the observation
when the count rate was higher (``high state'') is overlaid on
that of the remaining $\sim$50 ks  segment (``low state'').
The spectrum is slightly  harder during the high state with
a hardness ratio H/S = 3.17 as compared to H/S = 2.84 in the low-state.
The Fe K emission line complex 
is detected in both the high state and low state time segments.
Similar to the EPIC pn spectrum, the ACIS-S spectrum extracted
during the first 35 ks high-state segment reveals a
weak  emission feature near 6.4 keV that is undoubtedly fluorescent Fe
emission  (Fig. 4-bottom). A similar feature was present in the
off-axis zero-order ACIS-S spectrum of RY Tau in 2003 (Audard et al. 2005).

\subsubsection{Discrete Temperature Models}

We  fitted the undispersed spectra with absorbed $apec$ optically thin plasma
models to obtain estimates of the absorption column density (N$_{\rm H}$), 
mean plasma temperature (kT), metallicity ($Z$),  X-ray flux (F$_{x}$),
and unabsorbed X-ray luminosity (L$_{x}$).
Separate fits were obtained for spectra extracted during flare intervals
and intervals which excluded obvious flares, as summarized in Table 3.
The 1T $apec$ models are overly simplistic in the sense that they
only provide a mean  temperature and don't give information
on how the plasma is distributed versus temperature. The
distribution of plasma versus temperature is better assessed
with differential emission measure models (Sec. 5.2.2).

Several conclusions can be reached by examining the
fit results in Table 3.
Despite the simplicity of the  1T $apec$ model, it gives a
surprisingly good fit of the EPIC pn ``quiescent'' spectrum,
with a mean temperature of kT = 4.35 keV and reduced
chi-squared value $\chi^{2}_{\nu}$ = 1.02. A fit of the
same spectrum with a two-temperature (2T) $apec$ model
yields very little further improvement as gauged
by the fit statistic (Table 3 Notes). This is a clear indication
that the X-ray emission measure is dominated by hot plasma.

Mean temperatures were higher during flare segments.
Very high temperatures occurred during the 25 ks
flare segment in the {\em XMM-Newton} observation.
Fitting the pn spectrum using combined data from both flare 
peaks gives  kT$_{flare}$ = 14.6 [11.3 - 17.9] keV.
If spectra extracted for each flare peak are fitted 
separately, the second peak during which the hard band
count rate reached a maximum yields a slightly higher
temperature, but flare temperature uncertainties are large. 

The X-ray luminosity is clearly variable. As Table 3
shows, the broad-band luminosity was highest during
the high-state in the first $\sim$35 ks of the first
{\em Chandra} observation (log L$_{x}$ = 31.16 ergs s$^{-1}$). 
The luminosity during the second {\em Chandra} exposure
(ObsId 17539)  and
the {\em XMM-Newton} observation were lower and 
comparable to the value log L$_{x}$ = 30.67 ergs s$^{-1}$ 
obtained in 2009 December by {\em Chandra} (SAG11).
Based on existing observations, the typical luminosity
of RY Tau excluding  flares is
log L$_{x}$ = 30.65 ($\pm$0.10) ergs s$^{-1}$.
This is at the high end of the range observed for
cTTS  in Taurus (Fig. 1 of Telleschi et al. 2007a),
as discussed further in Sec. 6.1.

All 1T $apec$ fits converge to subsolar metallicity with
values in the range $Z$ $\approx$ 0.2 $Z_{\odot}$ (EPIC pn) to
$Z$ $\approx$ 0.4 $Z_{\odot}$ (ACIS-S). Fits in which the abundances
of individual elements were allowed to vary  show that 
the low $Z$ values are driven by  a low Fe abundance.
A low Fe abundance is also obtained using differential
emission measure models (Sec. 5.2.2).

X-ray absorption is best-determined from the EPIC pn spectra
which provide better sensitivity at low energies $\ltsimeq$1 keV
where absorption becomes important. The 1T $apec$ EPIC pn fit of 
the spectrum  with the flare interval excluded (Table 3) gives a
best-fit absorption 
N$_{\rm H}$ = 4.3 [4.1 - 4.4; 90\% conf.] $\times$ 10$^{21}$ cm$^{-2}$.
A fit of the pn spectrum during the flare interval gives a nearly 
identical N$_{\rm H}$ value (Table 3).
This above N$_{\rm H}$ is in good agreement with estimates
based on optically-determined A$_{\rm V}$ = 2.2 $\pm$ 0.2 mag
(Calvet et al. 2004) using the conversion 
N$_{\rm H}$ = 2.2$\times$ 10$^{21}$A$_{\rm V}$~cm$^{-2}$ (Gorenstein 1975),
which gives N$_{\rm H}$ = 4.8$\pm$0.4 $\times$ 10$^{21}$cm$^{-2}$.
By comparison, the conversion N$_{\rm H}$ = 1.6 $\times$ 10$^{21}$A$_{\rm V}$~cm$^{-2}$
of Vuong et al. (2003),
gives N$_{\rm H}$ = 3.5$\pm$0.3 $\times$ 10$^{21}$cm$^{-2}$,
slightly less than inferred from the EPIC pn fits.
However, the value inferred for N$_{\rm H}$ is somewhat 
sensitive to the spectral model used and more sophisticated
differential emission measure (DEM) models that allow
for a range of plasma temperatures  give 
somewhat larger  N$_{\rm H}$  values,
as discussed below.

\begin{deluxetable}{lccccc}
\tabletypesize{\scriptsize}
\tablewidth{0pc}
\tablecaption{RY Tau Discrete Temperature Model Spectral Fits (0-order)
   \label{tbl-1}}
\tablehead{
\colhead{Parameter}      &
\colhead{Value  }
}
\startdata
Telescope                               & CXO                & CXO                & CXO                & XMM (pn)                & XMM (pn)\tablenotemark{a}   \nl
ObsId                                   & 15722              & 15722              & 17539              & 722320101               & 722320101  \nl
Obs start date                          & 22 Oct. 2014       & 22 Oct. 2014       & 23 Oct. 2014       & 21 Aug. 2013            & 21 Aug. 2013 \\  
Time interval (ks)                      & 0.0 - 35.0         & 35.0 - 85.5        & 0.0 - 23.9         & 20.0 - 45.0             & 0.0 - 20.0, 45.0 - 90.0    \nl
Duration (ks)                           & 35.0               & 50.5               & 23.9               & 25.0                    & 65.0        \nl
State                                   & high-decay         & low-rise           & low-decay          & flares                  & non-flare       \nl
Model                                   & 1T $apec$          & 1T $apec$          & 1T $apec$          & 1T $apec$           & 1T $apec$\tablenotemark{b}        \nl
N$_{\rm H}$ (10$^{21}$ cm$^{-2}$)       & 5.1 [4.7 - 6.0]    & 6.2 [5.3 - 7.5]    & 4.6 [3.0 - 6.8]    & 4.0 [3.7 - 4.3]     & 4.3 [4.1 - 4.4]        \nl
kT$_{1}$ (keV)                          & 7.66 [6.56 - 9.70] & 4.56 [3.92 - 5.27] & 2.88 [2.37 - 3.55] & 14.6 [11.3 - 17.9]\tablenotemark{b}  & 4.35 [4.18 - 4.59]     \nl
norm$_{1}$ (10$^{-3}$ cm$^{-5}$)\tablenotemark{c}       & 3.69 [3.45 - 3.92] & 2.50 [2.26 - 2.80] & 1.57 [1.29 - 1.95] & 1.10 [1.07 - 1.13]  & 1.04 [1.00 - 1.07]      \nl
Abundances ($Z/Z_{\odot}$)              & \{0.4\}            & 0.4 [0.26 - 0.52]      & \{0.4\}          & \{0.2\}             & 0.2 [0.16 - 0.27]             \nl
$\chi^2$/dof                            & 138.5/116          & 132.1/98           & 28.9/24              & 389.6/262           & 487.6/479            \nl
$\chi^2_{\nu}$                          & 1.19               & 1.35               & 1.20                 & 1.49                & 1.02            \nl
Net counts (cts)                        & 2755               & 2326               & 578                  & 6132                & 13118             \nl
F$_{\rm X}$ (10$^{-12}$ ergs cm$^{-2}$ s$^{-1}$)          & 4.89 (6.70)           & 2.54 (3.94) & 1.27 (2.10)  & 1.60 (2.07)   & 1.03 (1.59)      \nl
log L$_{\rm X}$ (ergs s$^{-1}$)                           & 31.16                 & 30.93       & 30.65  & 30.65              & 30.53            \nl
\enddata
\tablecomments{
Based on  XSPEC (vers. 12.8.2) fits of the background-subtracted ACIS-S and EPIC pn spectra binned
to a minimum of 20 counts per bin.
The tabulated spectral parameters
are absorption column density (N$_{\rm H}$), plasma energy (kT),
and XSPEC component normalization (norm).
Abundances are referenced to  Anders \& Grevesse (1989).
Square brackets enclose 90\% confidence intervals.
Quantities enclosed in curly braces were held fixed during fitting.
The total X-ray flux (F$_{\rm X}$) 
is the absorbed value in the 0.2 - 10 keV range, followed in
parentheses by  unabsorbed value.
The total X-ray luminosity L$_{\rm X}$  is the  unabsorbed
value in the 0.2 - 10 keV range and  assumes a
distance of 134 pc.}
\tablenotetext{a}{If the MOS1 and MOS2 spectra are also included in the fit, similar results are 
obtained: N$_{\rm H}$ = 4.6 [4.4 - 4.7]e21 cm$^{-2}$, kT$_{1}$ = 4.21 [4.06 - 4.36] keV,
$Z/Z_{\odot}$ = 0.24 [0.20 - 0.29], norm$_{1}$ = 1.07 [1.03 - 1.10]e-03 cm$^{-5}$,
$\chi^2$/dof = 1176.5/845, $\chi^2_{\nu}$ = 1.39, 
F$_{\rm X}$  = 1.04 (1.63) $\times$ 10$^{-12}$ ergs cm$^{-2}$ s$^{-1}$. }
\tablenotetext{b}{Adding a second temperature component and allowing the abundances
of individual elements to vary (2T $vapec$ model) gives  
N$_{\rm H}$ = 5.2 [4.6 - 5.9]e21 cm$^{-2}$, kT$_{1}$ = 0.41 [0.33 - 0.81] keV, kT$_{2}$ = 4.29 [4.04 - 4.54] keV,
norm$_{1}$ = 0.09 [0.03 - 0.19]e$-$3 cm$^{-5}$, norm$_{2}$ = 0.89 [0.84 - 0.94]e$-$3 cm$^{-5}$, 
$\chi^2$/dof = 459.9/473,  $\chi^2_{\nu}$ = 0.97, F$_{\rm X}$ = 1.04 [1.78] ergs cm$^{-2}$ s$^{-1}$, log L$_{\rm X}$ = 30.58.
Element abundances relative to their solar values are Ne = 2.02 [1.30 - 3.31], Mg = 1.84 [1.00 - 2.82],
Si = 0.61 [0.14 - 1.12], Ca = 2.11 [0.65 - 3.59], Fe = 0.26 [0.20 - 0.33].
}
\tablenotetext{b}{The value of kT is an average during the flares. Separate fits of the two flare peaks
                  give different kT values.}
\tablenotetext{c}{For thermal $vapec$ models, the norm is related to the volume emission measure
                  (EM = n$_{e}^{2}$V)  by
                  EM = 4$\pi$10$^{14}$d$_{cm}^2$$\times$norm, where d$_{cm}$ is the stellar
                  distance in cm. At d = 134 pc this becomes
                  EM = 2.15$\times$10$^{56}$ $\times$ norm (cm$^{-3}$). }
\end{deluxetable}

\clearpage

\subsubsection{Differential Emission Measure Models}

The 1T $apec$ models converge to high plasma temperatures 
T $\sim$ 50 MK (kT $\sim$ 4 - 5 keV) so the X-ray emission is undoubtedly dominated 
by very hot plasma. However, the presence of low-temperature emission
lines in the grating spectra (Sec. 5.3) indicate that some
cool plasma (T $\ltsimeq$ 10 MK) is also present.
Differential emission measure (DEM)  models allow for plasma 
distributed over a range of temperatures and thus
provide a more realistic picture of the emission measure
distribution than discrete-temperature models.

We have reconstructed the DEM using the variable abundance XSPEC model
$c6pvmkl$ which is based on Chebyshev polynomials
\footnote{https://heasarc.gsfc.nasa.gov/xanadu/xspec/manual/XSmodelC6mekl.html}. 
We fitted the EPIC pn spectrum with this model, as well as a simultaneous
fit of the EPIC pn spectrum plus the RGS1 and RGS2 spectra. 
Flare intervals were excluded in the interest of obtaining
a picture of how the plasma is distributed during  
non-flaring (``quiescent'') conditions. Abundances of key elements with
detected emission lines in the grating spectra were allowed to vary.
Fit results are given in Table 4 and the derived DEM
distribution  from the best-fit model is shown in Figure 5-top.
There are only minor differences between the DEM fit results 
obtained fitting the EPIC pn spectrum alone and by fitting 
the pn and RGS spectra simultaneously.  This is a result of the
much higher signal-to-noise ratio in the pn spectrum, which
dominates the fit.

The DEM is dominated by a prominent peak
at kT $\approx$ 4 - 5 keV as expected from the 1T $apec$  models.
The high-temperature component is well-constrained
by the continuum and several highly-ionized Fe lines.
The DEM drops to a broad minimum near 1 keV (T $\sim$ 10 MK) and then rises
slowly toward lower energies. The shape of the DEM below
1 keV is not tightly constrained because it is quite sensitive to
absorption (N$_{\rm H}$) which becomes important at low energies
and suppresses low-energy emission.
Figure 5-bottom illustrates how relatively small changes
in absorption affect the derived shape of the cool plasma emission
measure distribution.  For the best-fit
absorption determined by the $c6pvmkl$ model (N$_{\rm H}$ =
6 $\times$ 10$^{21}$ cm$^{-2}$), the cool component rises 
steeply below 0.5 keV toward lower energies. But as N$_{\rm H}$ is decreased 
the cool component contributes less and becomes 
negligible if N$_{\rm H}$ = 4 $\times$ 10$^{21}$ cm$^{-2}$.
In addition to the sensitivity of the derived DEM to N$_{\rm H}$,
it is  worth keeping in mind
that  the inverse modeling which underlies  DEM reconstruction 
methods suffers from non-uniqueness issues, as has been 
discussed in  previous studies (e.g. Judge \& McIntosh 1999).
Thus, the DEM reconstruction shown in Figure 5 should not
be construed as unique.

The  absorption determined from the variable-abundance DEM fit of the
pn$+$RGS ``quiescent'' spectra
is N$_{\rm H}$ = 6.0 [5.2 - 7.1; 90\% conf.] $\times$ 10$^{21}$ cm$^{-2}$,
nearly identical to that obtained from 1T $apec$ fits of the ACIS-S 
low-state spectrum for ObsId 15722  (Table 3). It is also consistent
with the value N$_{\rm H}$ = 5.5 [4.7 - 6.8] $\times$ 10$^{21}$ cm$^{-2}$ 
obtained by {\em Chandra} ACIS-S in 2009 (SAG11) and in
the 2003 {\em Chandra} off-axis exposure (Audard et al. 2005).
The above absorption determinations are  consistent 
with the value expected based on A$_{\rm V}$ = 2.2 $\pm$ 0.2 mag using
the Gorenstein (1975) conversion which, as noted in Sec. 5.2.1, 
gives N$_{\rm H}$ = 4.8$\pm$0.4 $\times$ 10$^{21}$cm$^{-2}$.
But they are somewhat higher than the value 
N$_{\rm H}$ = 3.5$\pm$0.3 $\times$ 10$^{21}$cm$^{-2}$ 
obtained using the Vuong et al. (2003) conversion.
Thus, some excess X-ray absorption above that expected
from A$_{\rm V}$ may be present.

The DEM model fit  converges to a subsolar Fe abundance
Fe = 0.31 [0.23 - 0.38; 90\% confidence interval] $\times$ solar.
The other elemental
abundances are not as tightly constrained but O is 
also subsolar and the abundance ratio Ne/Fe $\approx$ 2.5 - 2.9  
is similar to values reported for other TTS in Taurus including the 
prototype T Tau (G\"{u}del et al. 2007b; Telleschi et al. 2007b).
The above abundances are relative to the solar reference values
of  Anders \& Grevesse (1989).

\begin{deluxetable}{lcc}
\tabletypesize{\scriptsize}
\tablewidth{0pc}
\tablecaption{RY Tau {\em XMM-Newton} DEM Model Spectral Fits
   \label{tbl-1}}
\tablehead{
\colhead{Parameter}      &
\colhead{Value  }
}
\startdata
Spectra                                             & pn                      & pn $+$ RGS1\&2          \nl
ObsId                                               & 722320101               & 722320101               \nl
Time interval                                       & quiescent               & quiescent               \nl
Model                                               & $c6pvmkl$               & $c6pvmkl$               \nl
N$_{\rm H}$ (10$^{21}$ cm$^{-2}$)                   & 6.0 [5.0 - 6.8]         & 6.0 [5.2 - 7.1]         \nl
norm$_{1}$ (10$^{-4}$ cm$^{-5}$)                    & 0.46 [0.22 - 0.87]      & 0.40 [0.20 - 0.84]  \nl
Abundances                                          & varied\tablenotemark{a} & varied\tablenotemark{b}                  \nl
$\chi^2$/dof                                        & 453.6/469               & 551.2/541                     \nl
$\chi^2_{\nu}$                                      & 0.97                    & 1.02                         \nl
F$_{\rm X}$ (10$^{-12}$ ergs cm$^{-2}$ s$^{-1}$)    & 1.04 (2.73)             & 1.04 (2.60) \nl
log L$_{\rm X}$ (ergs s$^{-1}$)                     & 30.77                   & 30.75       \nl
\enddata
\tablecomments{
Based on  XSPEC (vers. 12.8.2) fits of the background-subtracted  EPIC pn and RGS1\&2 spectra 
using a variable-abundance $c6pvmkl$ model.
Abundances are referenced to  Anders \& Grevesse (1989).
Square brackets enclose 90\% confidence intervals.
The total X-ray flux (F$_{\rm X}$) 
is the absorbed value in the 0.2 - 10 keV range, followed in
parentheses by  unabsorbed value.
The total X-ray luminosity L$_{\rm X}$  is the  unabsorbed
value in the 0.2 - 10 keV range and  assumes a
distance of 134 pc.}
\tablenotetext{a}{Abundances and 90\% confidence intervals are: O = 0.25 [0.05 - 0.55], 
 Ne = 0.72 [0.32 - 1.40], Mg = 1.19 [0.65 - 1.86], Si = 0.49 [0.16 - 0.82],
 Ca = 2.39 [0.98 - 3.84], Fe = 0.29 [0.22 - 0.37] $\times$ solar.}
\tablenotetext{b}{Abundances and 90\% confidence intervals are: O = 0.30 [0.10 - 0.67],
 Ne = 0.89 [0.37 - 1.47], Mg = 1.29 [0.72 - 1.87], Si = 0.52 [0.18 - 0.87], 
 Ca = 2.39 [0.95 - 3.88], Fe = 0.31 [0.23 - 0.38] $\times$ solar.}
\end{deluxetable}

\clearpage

\subsection{X-ray Grating Spectra}

\subsubsection{XMM-Newton RGS}

The first order RGS1 and RGS2 spectra for the full usable exposure
are shown in Figure 6. 
Emission lines and line photon fluxes are listed in 
Table 5, along with upper limits for important non-detections.
Line fluxes were measured by fitting lightly-binned  spectra 
using a Gaussian line profile fixed at the instrumental
width  and a power-law  model of the adjacent 
continuum. For faint lines the Gaussian centroid was
fixed at the line reference energy but for brighter
lines the centroid was allowed to vary.
Since the background is high and the line signal-to-noise
ratios are low, we analyzed total RGS1 and RGS2 spectra (source $+$ background)
as recommended by {\em XMM-Newton} RGS analysis 
guidelines\footnote{http://xmm.esac.esa.int/sas/current/documentation/threads/rgs\_thread.shtml}.

Visible lines at higher energies above 1 keV are  Ne X at
1.022 keV (12.134 \AA) and Mg XII at 1.47 keV (8.42 \AA). 
The Si XIII triplet near 1.86 keV may also be present but
is more clearly detected in the {\em Chandra} MEG spectrum (Fig. 7).
At low energies below 1 keV background begins to dominate and line
identifications become more uncertain. Fe XVII is present and 
the O VIII line at 0.654 keV (18.97 \AA) is  also visible in RGS2 
but is weak or absent in RGS1. 
The Ne IX triplet is not detected by RGS2 and there
is no RGS1 coverage of Ne IX. 

There is a noticeable feature in RGS1 at 
E = 0.577 $\pm$ 0.002 (1$\sigma$)  keV 
which is slightly higher than  the reference energy of the O VII
resonance  line (E$_{ref}$ = 0.574 keV). RGS2 lacks coverage
at this energy. There is a Ca XVI transition at E$_{ref}$ = 0.577 keV 
but RGS1 simulations do not reproduce the Ca line. If the feature
is interpreted as O VII then the rather large 3 eV blueshift is
not readily explained as a calibration 
offset\footnote{Details on RGS wavelength calibration can be found at \\
http://xmm2.esac.esa.int/docs/documents/CAL-TN-0030.pdf .} 
or a Doppler shift such as might arise if O VII formed  in the blueshifted jet. 
The latter interpretation would require  much higher jet speeds than have been 
obtained from  optical measurements (Sec. 6.4).  The identification of this 
feature as O VII is thus questionable and inspection of the RGS1 background
spectrum raises suspicions that it is noise-related.

\subsubsection{Chandra HETG}

The 1st order MEG spectrum from the longer first observation
(ObsId 15722) is shown in Figure 7 along with
a portion of the HEG spectrum in the vicinity of the 
high-temperature Fe K$\alpha$ line complex.  Several emission lines
are detected (Table 5) superimposed on a hot continuum.
The emission lines span a broad range of temperature as 
judged from maximum line emissivity temperatures of
$\sim$6 MK (Ne X)  up to $\sim$63 MK (Fe XXV;  
visible in HEG). There are no high-confidence HETG line detections at  
energies below 1 keV but weak Ne IX emission may be present
near 0.922 keV (13.447 \AA~) as discussed further in
Sec. 5.4. The absence of strong line detections below 1 keV
is attributable to the rapid falloff in ACIS-S/HETG effective
area at low energies and the effects of source absorption. 
We also note that potentially useful Ly$\beta$ lines from O and Ne 
are not detected so estimates of N$_{\rm H}$ using Ly$\alpha$/Ly$\beta$ 
flux ratios are precluded.

The excellent energy resolution and calibration of HETG 
permit a rigorous comparison of observed line centroid energies 
with their reference  (rest-frame) values and checks for line-broadening.
Best-fit first-order MEG line-centroid energies (Table 5) show no significant offsets from the
reference energies given in ATOMDB v3.0.2\footnote{www.atomdb.org}.  
Gaussian fits of the brightest
lines yield centroid energies that differ by at most $\pm$1 eV from reference energies,
well within MEG calibration accuracy\footnote{http://cxc.harvard.edu/proposer/POG/html/}.
Measured line-widths of the brightest lines (Si XIV, Si XIII$r$, Mg XII, Ne X) do not
exceed instrumental values. Thus, no evidence for significant centroid shifts or
excess line broadening is found for those lines bright enough to confidently 
measure line properties. 

The high plasma temperatures inferred from zero-order ACIS-S fits
are confirmed by continuum fits of the MEG spectrum using
intervals with no discernible line emission, as determined by
visual inspection and the ATOMDB line list. Fits of the MEG spectrum 
obtained during the first exposure (ObsId 15722) using
an absorbed 1T bremsstrahlung model give kT = 8.0 ($+$...,$-$4.3) keV
for the high-state interval and kT = 4.4 ($+$8.3,$-$1.9) keV for the
subsequent low-state, where the errors are 1$\sigma$. No useful constraint
on the upper limit of kT was obtained for the high-state spectrum.
Because of the limited number of channels
fitted after removing line intervals, no significant improvement
was obtained using a 2T  bremsstrahlung model.
The above values are similar to those obtained
from ACIS-S fits using 1T {\em apec} models (Table 3).

The 1st order MEG spectrum from the second shorter observation (ObsId 17539)
contains only 495 counts as compared to 3796 MEG counts in the first observation.
The lower number of counts in the second observation is due to the shorter
exposure time  and the lower count rate (Fig. 2). Since the spectrum 
was evolving between the first and second observations we have chosen not to
coadd the two MEG spectra. Only the brightest emission lines were detected in
the second observation (i.e. Si XIV Ly$\alpha$, Si XIII, Mg XII Ly$\alpha$).
The fluxes of these brightest lines are similar in the two observations and
their respective 1$\sigma$ confidence flux ranges overlap, albeit with
larger flux uncertainties in the second observation due to fewer  counts.
For the brightest line in the spectrum, Si XIV Ly$\alpha$, the flux from
the second observation is
F$_{\rm Si XIV}$ = 8.26 $\pm$ 3.37 (1$\sigma$) $\times$ 10$^{-6}$ ph cm$^{-2}$ s$^{-1}$,
in good agreement with the first observation (Table 5).

\begin{deluxetable}{llllclc}
\tabletypesize{\scriptsize}
\tablewidth{0pt}
\tablecaption{RY Tau Emission Line Properties\tablenotemark{a} }
\tablehead{
\colhead{Ion}      &
\colhead{$\lambda_{ref}$}     &
\colhead{$E_{ref}$}    & 
\colhead{$E_{obs}$}   &
\colhead{Line Flux (MEG)}   &
\colhead{Line Flux (RGS)}   &
\colhead{log $T_{e,max}$}  \\
\colhead{   }    &
\colhead{(\AA)     }    &
\colhead{(keV)     }    &
\colhead{(keV)  }   &
\colhead{(10$^{-6}$ ph cm$^{-2}$ s$^{-1}$)       }   &
\colhead{(10$^{-6}$ ph cm$^{-2}$ s$^{-1}$)       }   &
\colhead{(K)}    
}
\startdata
Fe XXV\tablenotemark{b,c} & 1.850   &  6.702     & 6.664     & 7.86 $\pm$ 4.52      & ...              & 7.8       \\
Ca XIX                    & 3.177   &  3.903     & 3.902     & 3.76 $\pm$ 1.56      & ...              & 7.5       \\
Si XIV Ly$\alpha$        & 6.182   &  2.006     & 2.006     & 8.34 $\pm$ 1.27      & ...              & 7.2       \\
Si XIII(r)                & 6.648   &  1.865     & 1.865     & 4.43 $\pm$ 1.02      & ...              & 7.0       \\
Si XIII(i)                & 6.688   & 1.854      & [1.854]   & 1.81 $\pm$ 0.80      & ...              & 7.0       \\
Si XIII(f)                & 6.740   &  1.840     & [1.840]   & 3.55 $\pm$ 0.91      & ...              & 7.0       \\
Mg XII                    & 7.106   &  1.745     & 1.744     & 1.73 $\pm$ 0.73      & ...              & 7.0       \\
Fe XXIV                   & 7.457   &  1.663     & 1.663     & 1.34 $\pm$ 0.65      & ...              & 7.3  \\
Mg XII Ly$\alpha$         & 8.421   &  1.472     & 1.472     & 5.54 $\pm$ 0.97 & 5.66 $\pm$ 3.51  & 7.0  \\
Mg XI(r)                  & 9.169   &  1.352     & 1.352     & 4.02 $\pm$ 1.57\tablenotemark{d}      & $\leq$3.36              & 6.8 \\
Mg XI(i)                  & 9.231   &  1.343     & [1.343]   & $\leq$1.14\tablenotemark{d}                   & ...      & 6.8        \\
Mg XI(f)                  & 9.314   &  1.331     & [1.331]   & $\leq$1.43\tablenotemark{d}                    & ...      & 6.8  \\
Fe XX\tablenotemark{b}    &10.021 &  1.237  & 1.237 & 1.95 $\pm$ 0.91 & ...     & 7.1 \\
Ne X Ly$\beta$            &10.239   &  1.211     & [1.211]   &$\leq$0.60             & ...            & 6.8   \\
Fe XXIV\tablenotemark{b}  &10.619   &  1.168     & 1.167\tablenotemark{e}        & 3.57 $\pm$ 1.12 & ...  & 7.3   \\
Fe XXIV                   &11.176   &  1.109     & 1.109 & 2.05 $\pm$ 0.97      & ...              & 7.3   \\
Fe XXIII                  &11.736   &  1.056     & 1.056 & 2.50 $\pm$ 1.24      & ...              & 7.2 \\
Ne X Ly$\alpha$           &12.134   &  1.022     & 1.021 & 3.99 $\pm$ 1.73      &  3.65 $\pm$ 1.72 & 6.8 \\
Ne IX(r)                  &13.447   &  0.922     & [0.922]   & $\leq$3.65\tablenotemark{g}    &$\leq$2.44     & 6.6 \\
Fe XVII                   &15.014   &  0.826     & [0.826]   & $\leq$2.19                    &  1.78 $\pm$ 1.46\tablenotemark{f,h}  & 6.8 \\
O VIII Ly$\beta$          &16.006   &  0.775     & [0.775]   & ...              &$\leq$3.05                & 6.5  \\
Fe XVII\tablenotemark{b}  &16.780   &  0.739     & 0.736 & ...                  & 1.59 $\pm$ 1.53\tablenotemark{f}  & 6.8 \\       
O VIII Ly$\alpha$         &18.969   &  0.654     &0.655  & ...                  & 2.44 $\pm$ 2.33\tablenotemark{f,i}  & 6.5 \\
O VII(r)                  &21.602   &  0.574     & [0.574]  & ...   & $\leq$3.09\tablenotemark{j}  & 6.3 \\
\enddata
\tablenotetext{a}{
Notes:~Observed line energies and fluxes are from continuum-subtracted 1st order HETG/MEG (ObsId 15722) 
and RGS (Obs 722320101) full-exposure spectra (flare$+$quiescent), unless otherwise noted.  
Tabulated quantities are: ion name (Ion) where r,i,f denote resonance, intercombination, and
forbidden lines of He-like triplets,  reference wavelength ($\lambda_{ref}$) and
energy  ($E_{ref}$) of transition from AtomDB version 3.0.2 (www.atomdb.org),
measured line centroid energy  ($E_{obs}$) where square brackets mean the value was held fixed during fitting, 
observed (absorbed) continuum-subtracted line flux (Line Flux) 
with 1$\sigma$ uncertainties (upper limits are 1$\sigma$), and maximum line emissivity electron temperature  ($T_{e,max}$).
An ellipsis means no reliable flux measurement was obtained.}
\tablenotetext{b}{Possible blend.}
\tablenotetext{c}{The observed line flux and centroid energy are from the 1st order HEG spectrum.
                  There are four closely-spaced Fe XXV lines in the range 6.637 - 6.702 keV (Fig. 7).
                  Their emissivity-weighted average energy is 6.682 keV.
                  The flux was measured with the Gaussian line width fixed at the instrumental value.
                  The observed feature is broadened, indicating that multiple lines contribute.}
\tablenotetext{d}{MEG fluxes and upper limits for Mg XI are from  high-state spectrum. Undetected in low-state.} 
\tablenotetext{e}{There is also weak emission from Fe XVIII/XX at 1.196 keV in the MEG spectrum.}
\tablenotetext{f}{Low significance feature; possible line detection.}
\tablenotetext{g}{Faint Ne IX emission may be present in MEG. See text.}
\tablenotetext{h}{Low-significance RGS features 
                  are also present in the Fe XVII complex at  0.727 - 0.739 keV.}
\tablenotetext{i}{The quoted O VIII line flux is from a  feature visible in RGS2. 
                  Not confirmed in RGS1 or MEG.}
\tablenotetext{j}{There is a feature visible at 0.577 $\pm$ 0.002 (1$\sigma$) keV  in RGS1
                  with observed flux 2.17 $\pm$ 2.06 ph cm$^{-2}$ s$^{-1}$.
                  There is no corresponding RGS2 coverage at this energy. Because of the slight
                  energy offset this feature cannot be conclusively identified as O VII$r$.} 
\end{deluxetable}

\clearpage

\begin{deluxetable}{lll}
\tabletypesize{\scriptsize}
\tablewidth{0pc}
\tablecaption{RY Tau He-like Triplet Emission Lines\tablenotemark{a} }
\tablehead{
\colhead{ } &
\colhead{Si XIII} &
\colhead{Mg XI }  \\
}
\startdata
Grating                      & HETG (MEG)             & HETG (MEG)            \nl
Time interval                & total                  & high-state            \nl
Flux($r$)\tablenotemark{b}   & 4.43 $\pm$ 1.02        & 4.02 $\pm$ 1.59       \nl
Flux($i$)\tablenotemark{b}   & 1.81 $\pm$ 0.80        & $\leq$1.14            \nl
flux($f$)\tablenotemark{b}   & 3.55 $\pm$ 0.91        & $\leq$1.43            \nl
G                            & 1.21$^{+0.86}_{-0.54}$ & $\leq$0.64            \nl
R                            & 1.96$^{+2.46}_{-0.95}$ & ...\tablenotemark{c}  \nl
R$_{0}$                      & 2.3                    & 2.7                   \nl
log T$_{e,max}$ (K)          & 7.0                    & 6.8                   \nl
log T$_{e}$ (K)              & 6.6$^{+0.4}_{-0.3}$    & $\geq$6.9             \nl
log n$_{e}$ (cm$^{-3}$)      & 13.0$^{+0.9}_{-...}$   & ...\tablenotemark{c}  \nl

\tablenotetext{a}{
Notes:~Line fluxes are from {\em Chandra} MEG 1st order spectra
($+$1 and $-$1 orders combined) for  ObsId 15722. The Mg XI
flux is from the high-state spectrum obtained during the first 
35 ks of the observation. 
The flux ratios are defined as G = (f $+$ i)/r and R = f/i. 
Uncertainties are $\pm$1$\sigma$. Flux upper limits are 1$\sigma$.
Limiting values for G and R 
use 1$\sigma$ flux upper limits.
R$_{0}$ is the theoretical low-density limit at maximum emissivity temperature 
T$_{e,max}$ (Porquet \& Dubau 2000; Porquet et al. 2001). The derived values of
T$_{e}$ and n$_{e}$ assume negligible photoexcitation.}
\tablenotetext{b}{Line photon flux in units of 10$^{-6}$ ph cm$^{-2}$ s$^{-1}$.} 
\tablenotetext{c}{No useful constraint obtained.}
\enddata
\end{deluxetable}

\clearpage

\subsection{He-like Triplets}

Line emission from  He-like triplets is summarized in  Table 6. 
Line flux ratios of the resonance (r), intercombination (i), and forbidden (f)
lines of He-like triplets are important plasma diagnostics (Porquet et al. 2001).
One or more of the
triplet lines of  Si XIII and  Mg XI is visible in the MEG first order spectrum.
Weak emission is also present in MEG  at the reference energies of the Ne IX $r$ 
and $i$ lines, as discusssed further below. The O VII triplet was not 
detected  by MEG and the identification of the feature in RGS1 
offset by 3 eV from the O VII$r$ reference energy is uncertain.

In He-like triplets, the line flux ratio $R$ = $f/i$ is sensitive to both
electron density $n_{e}$ and the UV radiation field (Gabriel \&  Jordan 1969).
If $\phi$ is the photoexcitation rate
from the $^{3}$S$_{1}$  level to the
$^{3}$P$_{2,1}$  levels, then
$R$ = $R_{0}$/[1 $+$ ($\phi$/$\phi_{c}$) $+$ ($n_{e}$/$n_{c}$)], where
$\phi_{c}$ and $n_{c}$ are the critical photoexcitation
rate and critical density. The quantities $R_{0}$, $\phi_{c}$,
and $n_{c}$ depend only on atomic parameters and the electron
temperature of the source, as defined in Blumenthal, Drake, \& Tucker (1972).
For cool stars it is usually assumed that photoexcitation is negligible
($\phi$/$\phi_{c}$ $\approx$ 0), as we do here. However, see Drake (2005)
for caveats regarding possible UV effects which can decrease the
$R$ ratio, mimicing high densities.
The ratio $G$ = ($f+i$)/$r$ is sensitive to electron temperature T$_{e}$
but is not strongly dependent on $n_{e}$.

The only He-like triplet for which the $r,f$ and $i$ lines were all
confidently detected is Si XIII in the MEG spectrum. Si XIII forms in 
hot plasma at a characteristic temperature T$_{e,max}$ $\sim$ 10 MK
and its $R$ ratio is sensitive only to high densities in the
range log $n_{e}$ $\approx$ 13 - 14 (Porquet \& Dubau 2000). 
The computed value $R$ = 1.96$^{+2.46}_{-0.95}$ based on MEG line fluxes 
is close to the low-density limit value $R_{0}$ = 2.3 (n$_{e}$ $<$ n$_{c}$), 
and the two values are consistent within the (large) 1$\sigma$ uncertainties.
The computed $R$ ratio formally gives log n$_{e}$ = 13.0$^{+0.9}_{-...}$
but since $R$ $\approx$ $R_{0}$ the lower bound on n$_{e}$ is 
essentially unconstrained and densities 
log n$_{e}$ $<$ 13.0 are allowed. 

Although the Mg XI$r$ line 
was detected in the MEG high-state spectrum, the $i$ and $f$ lines were not. 
Thus, $R$ is unconstrained for Mg XI but the upper limit on $G$ provides 
a lower bound log  T$_{e}$ $\geq$ 6.9 (K) which is slightly above the  
maximum line emissivity temperature  log  T$_{e,max}$ = 6.8 (K).  
The high value of log  T$_{e}$ inferred for Mg XI and the fact
that it was only detected by MEG during the high-state (i.e. during 
the first $\sim$35 ks of ObsId 15722) indicate  that it is
associated with hot plasma present in the high-state
and is  unlikely to be shock-related.

The Ne IX triplet is notably absent from the RGS spectrum but
a few counts are detected in the lower-background MEG spectrum
at the Ne IX$r$ and $i$ reference energies (Fig. 7-top). However, 
all MEG fits return zero
net line flux at the reference energy of Ne IX$f$ so the 
$f$ line is definitely not detected. The weak Ne IX 
emission in MEG may contain contributions from Fe XIX and Fe XXI and 
in any case the emission is too faint to obtain reliable line flux 
measurements. Thus, we only give an upper limit for Ne IX in Table 5.

\subsection{Summary of Spectral Analysis}

The X-ray spectrum of RY Tau reveals spectral lines from 
a broad range of plasma temperature  superimposed on a hot continuum.
The X-ray spectrum and light curves are highly variable,
including short-duration ($\sim$hours) high-temperature flares signaling 
strong magnetic activity superimposed on slower light curve modulation
spanning at least $\sim$one day.
The differential emission measure is dominated by very hot plasma with a 
peak near kT $\approx$ 4 - 5 keV (T $\approx$ 50 MK) and a weaker 
contribution from  cool plasma below 1 keV (T $\ltsimeq$ 10 MK). 
Higher  temperatures kT $\sim$ 8 keV (T $\sim$ 90 MK) are inferred during flares.
The unabsorbed  X-ray luminosity is variable with a  typical value 
log L$_{x}$ =  30.65 ($\pm$ 0.1) ergs s$^{-1}$ outside of flares.
The  absorption column density N$_{\rm H}$ is comparable to 
or perhaps slightly greater than that predicted from 
A$_{\rm V}$. The Fe abundance is significantly subsolar. 
Undispersed CCD spectra show a faint emission
line  near 6.4 keV from fluorescent Fe arising in cold
material near the star irradiated by the hard X-ray source.
No significant line centroid shifts or line broadening
were detected for the brightest lines in {\em Chandra}
MEG spectra. The  $R$ ratio computed for the Si XIII He-like
triplet is consistent with that expected in the low-density limit.
The Mg XI triplet resonance line is only visible 
in the {\em Chandra} MEG spectrum during the high-state
when  a high plasma temperature was inferred and is 
thus evidently associated with hot plasma, not shocks.
Low-temperature emission lines are generally faint or
absent as a result of absorption and RGS noise, although
Fe XVII (T$_{e,max}$ $\sim$ 6 MK) and O VIII Ly$\alpha$ 
(T$_{e,max}$ $\sim$ 3 MK) are visible in the RGS spectra.

\vspace*{0.5in}

\section{Discussion}

\subsection{RY Tau in Context:  the Taurus CTTS Population}

RY Tau is one of the most massive and rapildly-accreting  cTTS 
in Taurus. It is remarkably similar to the prototype T Tau N
which dominates the X-ray and optical emission of the multiple T Tau
system (G\"{u}del et al. 2007b). RY Tau and T Tau N have
similar masses, accretion rates, N$_{\rm H}$, and Ne/Fe abundance
ratios (Calvet et al. 2004; G\"{u}del et al. 2007b). Both stars
have X-ray spectra consisting of an admixture of very 
hot and very cool plasma with  both components exhibiting slow
light curve variability (Fig. 2 of G\"{u}del et al. 2007b). 
The  X-ray luminosity of T Tau is 
log L$_{x}$(0.3 - 10 keV) = 31.18 ergs s$^{-1}$ (d = 140 kpc),
comparable to or slightly greater than that of RY Tau.
However, there are a few notable differences.
T Tau is a multiple system consisting of at least three 
closely-spaced objects (T Tau N, T Tau Sa,b) whereas binarity in
RY Tau is suspected but has so far not been proven.
In addition, no well-collimated large-scale optical jet such as that 
observed for RY Tau has yet been reported for the T Tau system.

A more quantitative comparison of RY Tau with other cTTS in
Taurus can be made using established correlations of 
L$_{x}$ with stellar mass (M$_{*}$) and  luminosity (L$_{*}$) 
identified in the XEST cTTS sample (Telleschi et al. 2007a).
The correlation with stellar mass is of the form
log L$_{x}$ = $A$$\cdot$log(M$_{*}$/M$_{\odot}$) $+$ $B$ (ergs s$^{-1}$)
where  $A$ = 1.70 $\pm$ 0.20 and $B$ = 30.13 $\pm$ 0.09 were determined
from the parametric estimation (EM)  method and $A$ = 1.98 $\pm$ 0.20 and 
$B$ = 30.24 $\pm$ 0.06 from the bisector method. We take L$_{x}$
as the known quantity derived from our X-ray spectral fits and
adopt log L$_{x}$ = 30.65 $\pm$ 0.1 ergs s$^{-1}$ (d = 134 pc)
as a typical value for RY Tau outside of flares, where the uncertainties
reflect only the range of low-state (``quiescent'') L$_{x}$ values 
(Tables 3 and 4). Adjusting this value upward to log L$_{x}$ = 30.69 $\pm$ 0.1 
ergs s$^{-1}$ at the distance of 140 pc used in the XEST
study gives  M$_{*}$ = 2.1 (1.8 - 2.7) M$_{\odot}$ from the EM-method
relation and M$_{*}$ = 1.7 (1.5 - 1.9) M$_{\odot}$ from the bisector method,
where the range in parentheses accounts only for the uncertainties in 
XEST regression fit parameters. These mass estimates are in very 
good agreement with previously published estimates for RY Tau (Table 1).
Thus, even though L$_{x}$ for RY Tau is among the highest of cTTS
studied in Taurus (Fig. 1 of Telleschi et al. 2007a), it is
consistent with expectations given that its mass is also high.

XEST regression fit results for the EM and bisector methods are similar
for the  L$_{x}$ versus L$_{*}$ correlation for cTTS in Taurus, which is
log L$_{x}$ = $C$$\cdot$log(L$_{*}$/L$_{\odot}$) $+$ $D$ (ergs s$^{-1}$)
where $C$ = 1.16 $\pm$ 0.09 and $D$ = 29.83 $\pm$ 0.06. Inserting
log L$_{x}$ = 30.69 for RY Tau yields L$_{*}$ = 5.5 (4.4 - 7.2) L$_{\odot}$.
This regression fit prediction is a  bit lower than previously
reported L$_{*}$ values (Table 1) but  the upper limit  L$_{*}$ = 7.2 L$_{\odot}$
is nearly equal to the value 7.6 L$_{\odot}$ determined by
Kenyon and Hartmann (1995). Given that the spectral type of
RY Tau is somewhat uncertain (and hence its  bolometric 
correction as well) and that the low-state (``quiescent'') L${_x}$ value is also 
uncertain by $\pm$0.1 dex,  the above difference is not  significant.
We conclude that published values of M$_{*}$ and L$_{*}$
for RY Tau are in acceptable agreement with predictions based on
XEST correlations for cTTS in Taurus.

\subsection{The 6.4 keV  Fluorescent Fe Line}

The 6.4 keV fluorescent Fe emission line is visible in both the
EPIC pn and ACIS-S flare or high-state spectra, but is not present in the
quiescent or low-state spectra (Figs. 3 and 4). Thus, the fluorescent line
is excited by hard X-rays produced in flares or high emission states
irradiating cold nearby material. Photon enegies E $>$ 7.11 keV
are needed to eject a K-shell electron which is followed
by a downward transition (e.g. from the L-shell) to produce the
6.4 keV line. The association of fluorescent
Fe emission with flares in young stars and protostars has
been seen before, but in  unusual cases such as the protostar
NGC 2071 IRS1 the 6.4 keV feature is present even in the absence
of discernible flares (Skinner et al. 2007; 2009).

Measurements of the continuum-subtracted 6.4 keV line flux  in the ACIS-S 
flare-segment spectrum (ObsId 15722) give
F$_{x,6.4}$ = 7.5 ($\pm$1.5) $\times$ 10$^{-14}$ ergs cm$^{-2}$ s$^{-1}$,
where the uncertainty is 90\% confidence.
The underlying continuum flux density is
F$_{x,cont}$ = 4.4 ($\pm$0.4) $\times$ 10$^{-13}$ ergs cm$^{-2}$ s$^{-1}$ keV$^{-1}$.
The above values give a line equivalent width EW = 0.17 $\pm$ 0.05 keV.
For comparison, the line flux from the 2003 off-axis ACIS-S spectrum  was
F$_{x,6.4}$ = 4.4 ($\pm$2.0) $\times$ 10$^{-14}$ ergs cm$^{-2}$ s$^{-1}$
(Audard et al. 2005). In our 2013 {\em XMM-Newton} observation the 
EPIC pn 6.4 keV line is narrow and weak and the line flux is 
quite uncertain but line flux estimates are comparable to that given 
above for the  2003 {\em Chandra} off-axis observation.

The fluorescent line equivalent width (EW) is related to
the column density of cold fluorescent material in the
optically thin slab approximation by
EW $\approx$ 2.3 N$_{24}$ [keV]  (Kallman 1995),
where N$_{24}$ is the column density of the cold matter
in units of 10$^{24}$ cm$^{-2}$. Using the value EW = 0.17 $\pm$ 0.05 keV
from the 2014 ACIS-S flare spectrum gives 
N$_{\rm H,cold}$ = 7.4$\pm$2.2 $\times$ 10$^{22}$ cm$^{-2}$.
This value is an order of magnitude greater  than inferred from
the  spectral fits (Table 3) so
dense cold target material located near the star but off the
line-of-sight is required. Accreting gas or disk gas
are two likely possibilities for the irradiated material.

\subsection{Slow Variability}

The slow decline followed by a slow rise  in the
{\em Chandra} light curves of the first observation  
is present in very soft, medium, and hard energy bands (Fig. 2).
This is a strong clue that the very soft, medium, and hard-band
emission have a common origin (Sec. 6.3).
Similar slow X-ray light curve variability
is present in the shorter second {\em Chandra} observation (Fig. 2)
and was previously seen in the 2009 {\em Chandra} observation (Fig. 2 of SAG11)
and the 2003 {\em Chandra} off-axis exposure.
Furthermore, the EPIC pn light curve in the 1 - 2 keV
band on 2013 August 21-22 is rising slowly but no slow
variability is obvious in the pn hard-band where rapid
flares are most conspicuous (Fig. 1).
Taken together, these results suggest that RY Tau 
is in a persistent state of slow X-ray modulation
accompanied by intermittent rapid flares.

In order to determine how the X-ray parameters changed 
with time during the slow decay and rise in the new {\em Chandra} observation
(ObsId 15722), we  extracted  ACIS-S spectra from 
five non-overlapping time intervals that span the 85 ks observation.
Each spectrum was fitted with an absorbed isothermal 1T $apec$ model
with metallicity fixed at $Z$ = 0.4 $Z_{\odot}$.
The variation of the mean plasma temperature  kT and absorbed broad-band 
X-ray flux F$_{x}$ versus time  are shown in Figure 8, along with the 
hard-band ACIS-S light curve.

It is apparent from Figure 8 that kT and F$_{x}$  declined
steadily until the  hard-band count rate reached a
minimum at elapsed time $t$ $\approx$ 40 - 50 ks and then they 
increased. The emission measure, as gauged by the XSPEC model $norm$,
mimics the time behavior of kT and F$_{x}$.
An exponential fit of the the kT versus time decay
profile (Fig. 8-top) gives an  e-folding time of 52.6 ks (14.6 hr).

What is intriguing about the time evolution is that even though 
kT clearly increased near the end  of the observation
after the hard-band count rate reached minimum, there is no 
clear signature of an impulsive flare in the hard-band
light curve (Fig. 2) that might have triggered the reheating.
Although a slowly-developing flare could be responsible
for the gradual brightening after the {\em Chandra} light curve 
reached minimum, the absence of any discernible  hard-band 
flare in combination with the similar slowly-changing count rates
in different energy bands is more suggestive of variability
associated with one or more surface features rotating across the 
line-of-sight. The conclusion  that the slow light curve variability may be 
linked to surface structures  rotating across the 
line-of-sight is tentative since no definite stable period of
order $\sim$days that could arise from stellar rotation
has yet been found in X-rays or optical. However, 
Holtzman et al. (1986) have presented compelling arguments
for surface structures (``spots'') on RY Tau based on 
their  analysis of optical photometric and spectroscopic
variability.

Although no definite rotation period for RY Tau has yet been
found, a rough estimate can be derived based on published
stellar parameters. Estimates of the stellar radius
range from R$_{*}$ = 2.9 $\pm$ 0.4  R$_{\odot}$ (Calvet et al. 2004)
to 5.0 $\pm$ 0.3  R$_{\odot}$ (Takami et al. 2013).
Petrov et al. (1999) obtained a projected rotational velocity 
$v$sin$i$ = 52 $\pm$ 2 km s$^{-1}$.
The disk inclination is quite uncertain with published estimates
ranging from $i_{disk}$ = 25$^{\circ}$ $\pm$ 3$^{\circ}$ (Akeson et al. 2005)
to $i_{disk}$ = 66$^{\circ}$ $\pm$ 2$^{\circ}$ (1.3 mm data) or 
71$^{\circ}$ $\pm$ 6$^{\circ}$ (2.8 mm data), the last two
being derived from CARMA mm interferometry (Isella et al. 2010).
Based on the above we adopt the representative values R$_{*}$ = 4  $\pm$ 1 R$_{\odot}$,
$v$sin$i$ = 52 $\pm$ 2 km s$^{-1}$ , and
$i_{disk}$ = 48$^{\circ}$ $\pm$ 24$^{\circ}$. Under the additional
assumption that the stellar and disk rotation axes are aligned
we obtain P$_{rot}$ =  2.9 (1.2 - 4.8) days where the range in
parentheses reflects the spread in  adopted stellar parameters.
We emphasize that the above is just an estimate and not a substitute for an
observational period determination.

We are not aware of any reports of optical periods of $<$5 days for
RY Tau. A slight peak of 5.6 d was noted in a periodogram by
Herbst et al. (1987) but after more detailed analysis no significant
power near 5.6 d was seen. Later attempts to recover the 5.6 d
signal also gave negative results (Herbst \& Koret 1988; Bouvier et al. 1993).
A low-significance peak of 7.5 d was claimed by Zajtseva (2010) 
in optical photometry obtained during 1996 but to our knowledge
this period has not been confirmed. There have been reports
of periods in the range $\sim$20 - 24 d but these are too long
to be due to rotation as was noted by Bouvier et al. (1993).

On a rapidly-rotating accreting young star like RY Tau
surface structures could originate in coronal active regions or at
accretion footpoints, giving rise to X-ray modulation.
Periodic X-ray modulation was found in 23 young stars
in the {\em Chandra} Orion COUP sample (Flaccomio et al. 2005).
In most cases the X-ray period was close to the known optical
period (P$_{opt}$ $\sim$2 - 14 d) but in some cases the 
X-ray period was about half the optical period. Compact sizes
less than (or much less than) R$_{*}$ were inferred for the X-ray
structures responsible for the modulation. In addition, possible
periodic X-ray modulation in the eruptive young star V1647 Ori
was attributed to surface structures associated with accretion
footpoints (Hamaguchi et al. 2012).

Since the slow {\em Chandra} light curve variability is present in very soft
and hard energy bands, it is unlikely that the surface 
structures thought to underlie the variability are accretion footpoints.
The predicted accretion shock temperature for RY Tau is
T$_{s}$ $\sim$ 3 MK (Sec. 6.4) and such cool plasma would
have little if any effect on the hard-band (2 - 8 keV) light
curve. Identification of the presumed surface structures
with one or more active regions spanning a range of temperature 
(and density) seems to be in better accord with the 
light curve variations.  Any  role that the hypothesized
(but as yet unseen) companion might play in the X-ray
variability cannot be reliably assessed without specific 
information regarding the companion type and its orbit. 

The  volume V of the X-ray emitting region can be estimated
from the volume emission measure EM = n$_{e}^{2}$V,
where n$_{e}$ is the average electron density in the emitting
region. A numerical value of EM is obtained from the
XSPEC $norm$ in spectral fits via the relation (Table 3 Notes)
EM = 2.15 $\times$ 10$^{56}$ $\times$ $norm$ (cm$^{-3}$).
In non-flaring states the 1T $apec$
fits give $norm$ $\approx$ (1 - 2.5) $\times$ 10$^{-3}$
cm$^{-5}$ (Table 3) and as a representative value we adopt 
$norm$ $\approx$ 1.5 $\times$ 10$^{-3}$ cm$^{-5}$.
This yields EM = 3.2 $\times$ 10$^{53}$ cm$^{-3}$ = n$_{e}^{2}$V
which would include any excess X-ray emission measure from 
active regions as well as a basal contribution from the
ambient corona. As such, it is only an upper limit on any 
active region  contribution. For a stellar radius 
R$_{*}$ = 2.9 R$_{\odot}$ (Calvet et al. 2004) the 
ratio of  emitting region volume to stellar volume is
V/V$_{*}$ $\approx$ 0.1/n$_{e,10}^2$ where
n$_{e,10}$ is the average density in units of 10$^{10}$ cm$^{-3}$.

Coronal densities in late-type stars determined from 
O VII and Ne IX triplet line ratios are typically
in the range n$_{e}$ $\sim$ 10$^{10}$ - 10$^{12}$ cm$^{-3}$
(Ness et al. 2004). Active region densities on the Sun
obtained from spatially-resolved {\em Hinode} 
observations are in the range 
n$_{e}$ $\sim$ 10$^{8}$ - 10$^{11}$ cm$^{-3}$ (Pradeep et al. 2012),
being at the high end of this range in the active region core.
For an assumed value n$_{e}$ $\sim$ 10$^{10}$ cm$^{-3}$ the relation
obtained above for RY Tau gives V $\approx$ 0.1V$_{*}$ 
and a characteristic emitting region radius
R = V$^{1/3}$ $\approx$ 0.75 R$_{*}$.
For a surface filling factor $f$ the relation R = $\sqrt{f}$R$_{*}$
yields $f$ = 0.56. Higher densities are certaintly possible 
as referenced above and are compatible with the Si XIII
triplet line ratios (Table 6), so a small emitting volume V $<$ V$_{*}$
seems assured and the  filling factor need not be large.

\subsection{Cool Plasma: Shocks or Corona?}

The presence of the O VIII Ly$\alpha$ emission line
in the RGS2 spectrum supports the conclusion from DEM
fits (Fig. 5) that some cool plasma is present.
The O VIII line has a maximum emissivity temperature 
T$_{e,max}$ $\approx$ 3 MK but the  emissivity
is still substantial up to higher temperatures of  $\sim$5 - 6 MK.  
Cool plasma at T $\ltsimeq$ 3 MK could potentially arise in
shocks, and a cool coronal component
at T $\sim$3 - 6 MK could also produce the O VIII line.

{\em Shocked Jet}:~Plasma temperatures of $\sim$3 MK are difficult to achieve
for a shocked jet in RY Tau based on current jet speed estimates. 
As discussed in SAG11, the maximum shock temperature for a shock-heated
jet is $T_{s}$ = 0.15$v_{s,100}^2$ MK, where $v_{s,100}$ is the  jet
speed in units of 100 km s$^{-1}$ (Raga et al. 2002). The optically-derived
jet speed for  RY Tau is $v_{jet}$ $\approx$ 165 km s$^{-1}$ which 
leads to a maximum predicted shock temperature
$T_{s}$ $\approx$ 0.4 MK (k$T_{s}$ $\approx$ 0.035 keV).
At this temperature, almost all of the X-ray emission would
emerge at energies below 0.2 keV where {\em XMM-Newton} or {\em Chandra}
have very little sensitivity. Unless the jet speed is higher than
estimated from optical observations or other jet-heating mechanisms
besides shocks are at work, a jet origin for the  cool X-ray
plasma  is difficult to justify.

{\em Accretion Shock}:~The case for producing cool (T $\sim$ 3 MK) X-ray 
plasma and the O VIII line by an  accretion shock is more favorable.
The post-shock temperature for a strong accretion shock is

\begin{equation}
T_{s} = 2.27 \times 10^{5}\mu\left[\frac{v_{s}}{100~\rm{km~s^{-1}}}\right]^{2}~K.
\end{equation}
where $\mu$ is the mean mass (amu) per particle in the accreting gas and $v_{s}$ is 
the free-fall speed at the  the shock interface  (Calvet \& Gullbring 1998)
To estimate $v_{s}$ we adopt the stellar parameteres of  Calvet et al. (2004), namely
M$_{*}$ = 2.0 M$_{\odot}$, R$_{*}$ = 2.9 R$_{\odot}$. The  accretion rate is
somewhat uncertain but the various models considered by Schegerer et al. (2008) are 
compatible with  rates of
$\dot{M}_{acc}$ $\approx$ (2.5 - 9.1) $\times$ 10$^{-8}$ M$_{\odot}$ yr$^{-1}$.
We adopt a value in the middle of this range $\dot{M}_{acc}$ = 
5 $\times$ 10$^{-8}$ M$_{\odot}$ yr$^{-1}$.
In the absence of a magnetic
field measurement we assume a typical cTTS value
B$_{*}$ $\approx$ 2000 G (Johns-Krull 2007). Using equation (1)
of K\"{o}nigl (1991) we obtain an inner disk truncation radius
from which the infalling material is assumed to originate
of $r_{in}$ = 7.15 R$_{*}$, assuming spherical accretion and 
H-ionized solar abundance plasma ($\mu$ = 0.6). Equation (1)
of Calvet \& Gullbring (1998) then gives $v_{s}$ = 479 km s$^{-1}$
corresponding to a post-shock temperature T$_{s}$ = 3.1 MK (kT$_{s}$ = 0.27 keV). 
The value of T$_{s}$ is only weakly-dependent on the
poorly-known values B$_{*}$ and $\dot{M}_{acc}$.

For our adopted stellar parameters the post-shock electron
density for a strong shock is (eq. [3] of Telleschi et al. (2007b)
$n_{2}$ $\approx$ 5.7 $\times$ 10$^{10}$${\dot{M}_{acc,-8}}/f$
cm$^{-3}$, where $\dot{M}_{acc,-8}$ is in units of 10$^{-8}$ M$_{\odot}$ yr$^{-1}$ and $f$
is the surface filling factor which is not well-known but is usually taken to be in
the range $f$ = 0.001 - 0.1.  For ${\dot{M}_{acc,-8}}$ = 5 and $f$ $\leq$ 0.1
the lower limit is $n_{2} \gtsimeq$ 2.8 $\times$ 10$^{12}$ cm$^{-3}$.
This value is higher than typical
coronal densities n$_{e}$ $\sim$ 10$^{10}$ - 10$^{11}$ cm$^{-3}$  
for active stars based on O VII triplet $R$ ratios (Ness et al. 2004).
The total accretion luminosity is (Calvet \& Gullbring 1998) 
$L_{acc}$ = 3.6 $\times$ 10$^{33}$ ergs s$^{-1}$ (= 0.94 L$_{\odot}$). 
For comparison, we note that the
observed (absorbed) flux of the O VIII Ly$\alpha$ line gives
L$_{\rm OVIII}$ = 4.8 $\times$ 10$^{27}$ ergs s$^{-1}$ with an uncertainty
of about a factor of two. The intrinsic (unabsorbed) luminosity depends
sensitively on N$_{\rm H}$ toward the line-forming region and would be
larger.

We have fitted the RGS2 spectrum (flares excluded) using 
a 2T $vapec$ optically thin plasma model to estimate the temperature
range needed to reproduce the O VIII Ly$\alpha$ line. Absorption was restricted  
to the range (4 - 6) $\times$ 10$^{21}$ cm$^{-2}$ and the hot
plasma component temperature was fixed at kT$_{hot}$ = 4.35 keV
as determined from EPIC pn fits. Subsolar
abundances were adopted with Fe = 0.3 and O = 0.3 $\times$ solar.
In order to obtain sufficient flux to reproduce the O VIII
line, cool component temperatures in the range 
kT$_{cool}$ $\approx$ 0.13 - 0.28 keV (T$_{cool}$ $\approx$ 1.5 - 3.2 MK)
are required. These values are consistent with the
accretion shock temperature estimated above. Even though
accretion shock temperatures are favorable for producing
O VIII, the fraction of soft X-rays that could escape and
be detected is sensitive to the line-of-sight absorption
toward the shock region and photon escape becomes problematic
at high infalling gas densities (Drake 2005).

{\em Cool Corona}:~Coronal plasma in active low-mass stars is distributed
over a wide range of temperature and can  include both a hot component
(T$_{hot}$ $\gtsimeq$ 10 MK) as well as  a cool component extending down to 
temperatures of T$_{cool}$ $\sim$ 3 - 4 MK. Two examples are the young
solar analog EK Dra (G\"{u}del et al. 1997) and the active binary
II Peg (Huenemoerder, Canizares, \& Schulz 2001), the latter showing
a strong coronal O VIII Ly$\alpha$ line.

Electron density information derived from O VII and Ne IX He-like 
triplets is typically used to discriminate
between cool dense plasma in an accretion shock and cool lower-density coronal
plasma. In RY Tau the O VII and Ne IX triplets are only weakly
detected if at all, so cool plasma density information is lacking.
However, as we have noted, the slow variability seen in the {\em Chandra} 
very-soft and medium  band light curves mimics that of the hard-band (Fig. 2),
providing compelling evidence  that some or all of the  very soft band emission
shares a common origin with the hard-band emission.
The hard-band emission  is undoubtedly associated with very hot 
plasma that cannot be reconciled with  cool shocks and is thus 
presumably coronal, but possible contributions from
the star-disk magnetic interaction region are not ruled out
at the existing limits of X-ray telescope spatial resolution.
By association, at least some of the cool plasma must also
be of magnetic (non-shock) origin.

\section{Summary}

We have presented new results clarifying  the X-ray
properties of the cTTS RT Tau based on observations 
obtained with {\em Chandra} and {\em XMM-Newton}. The main
results of this study are the following:

\begin{enumerate}

\item The X-ray emission of RY Tau is strongly variable,
consisting of intermittent rapid flares typical of coronal
magnetic activity  superimposed on slow light-curve modulation that
may be tied to rotation of surface features across the
line-of-sight.

\item The absorption column density 
N$_{\rm H}$ $\approx$ (4 - 6) $\times$ 10$^{21}$ cm$^{-2}$ determined 
from X-ray spectral fits is comparable to or slightly larger than
anticipated from  A$_{\rm V}$.

\item The characteristic X-ray luminosity of RY Tau 
log L$_{x}$ $\approx$ 30.65 ergs s$^{-1}$ is among the highest of
cTTS in Taurus but is nevertheless consistent with expectations
based on its rather high mass and a known correlation between
L$_{x}$ and stellar mass in the Taurus cTTS population.

\item The X-ray emission measure distribution of RY Tau is
dominated by hot plasma at characteristic temperatures of
kT$_{hot}$ $\sim$ 4 - 5 keV (T$_{hot}$ $\sim$ 50 MK),
but higher temperatures are recorded during flares.
Flares give rise to fluorescent emission from neutral or
near-neutral Fe at 6.4 keV arising from irradiated cold dense 
gas near the star. Gas in the accretion disk or accretion
streams provides potential fluorescent target material.

\item A cool plasma component is present which varies slowly
in lockstep with hotter plasma, providing a strong clue that at least some
of the cool plasma is  physically associated with the hotter plasma.
Shocks cannot explain the very high temperatures of hot plasma
which  is undoubtedly associated with magnetic heating processes in
the corona or perhaps in the star-disk magnetic interaction region.
By association, at least some of the cool plasma is also of 
magnetic (non-shock) origin.  

\item Any contribution to the  X-ray emission measure  from 
cool  plasma (T$_{cool}$ $\ltsimeq$ 3 MK) originating in the shocked
jet or an accretion shock is small compared
to the dominant hot plasma, but an accretion shock origin for 
the O VIII Ly$\alpha$ line is not ruled out.

\end{enumerate}

\acknowledgments

This work was supported by {\em Chandra} award GO4-15012X issued by the 
Chandra X-ray Observatory Center (CXC) and by NASA ADAP award NNX14AJ60G.
The CXC is operated by the 
Smithsonian Astrophysical Observatory (SAO) for and on behalf of 
NASA under contract NAS8-03060. 
This work is partially based on observations obtained with 
{XMM-Newton}, an ESA science mission with instruments and
contributions directly funded by ESA member states and
the USA (NASA). This research has made use of the HEASOFT
data analysis software develeoped and maintained by HEASARC
at NASA GSFC.

\clearpage


\begin{figure}
\figurenum{1}
\epsscale{1.0}
\includegraphics*[width=7.4cm,height=10.0cm,angle=-90]{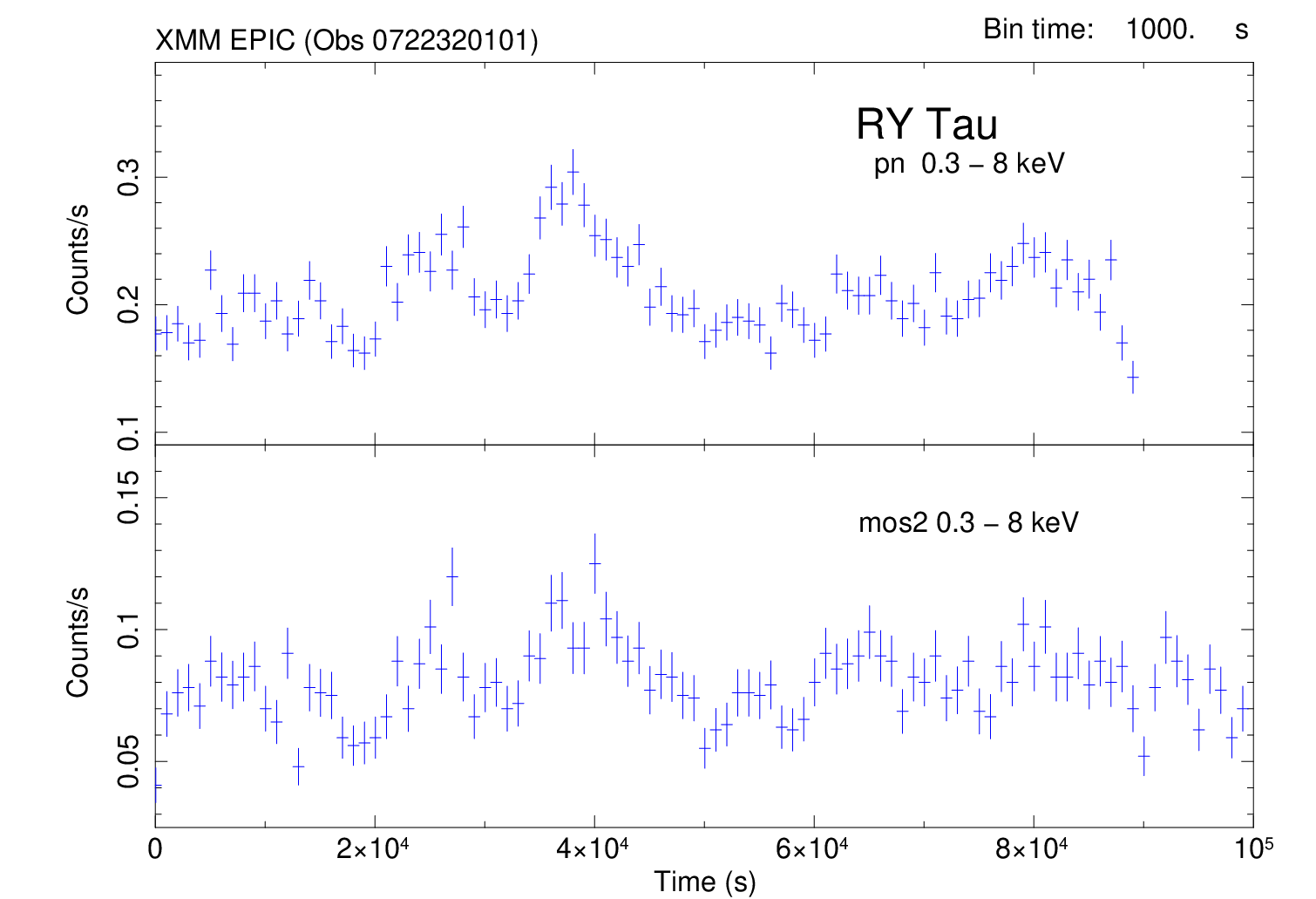} \\
\includegraphics*[width=7.4cm,height=10.0cm,angle=-90]{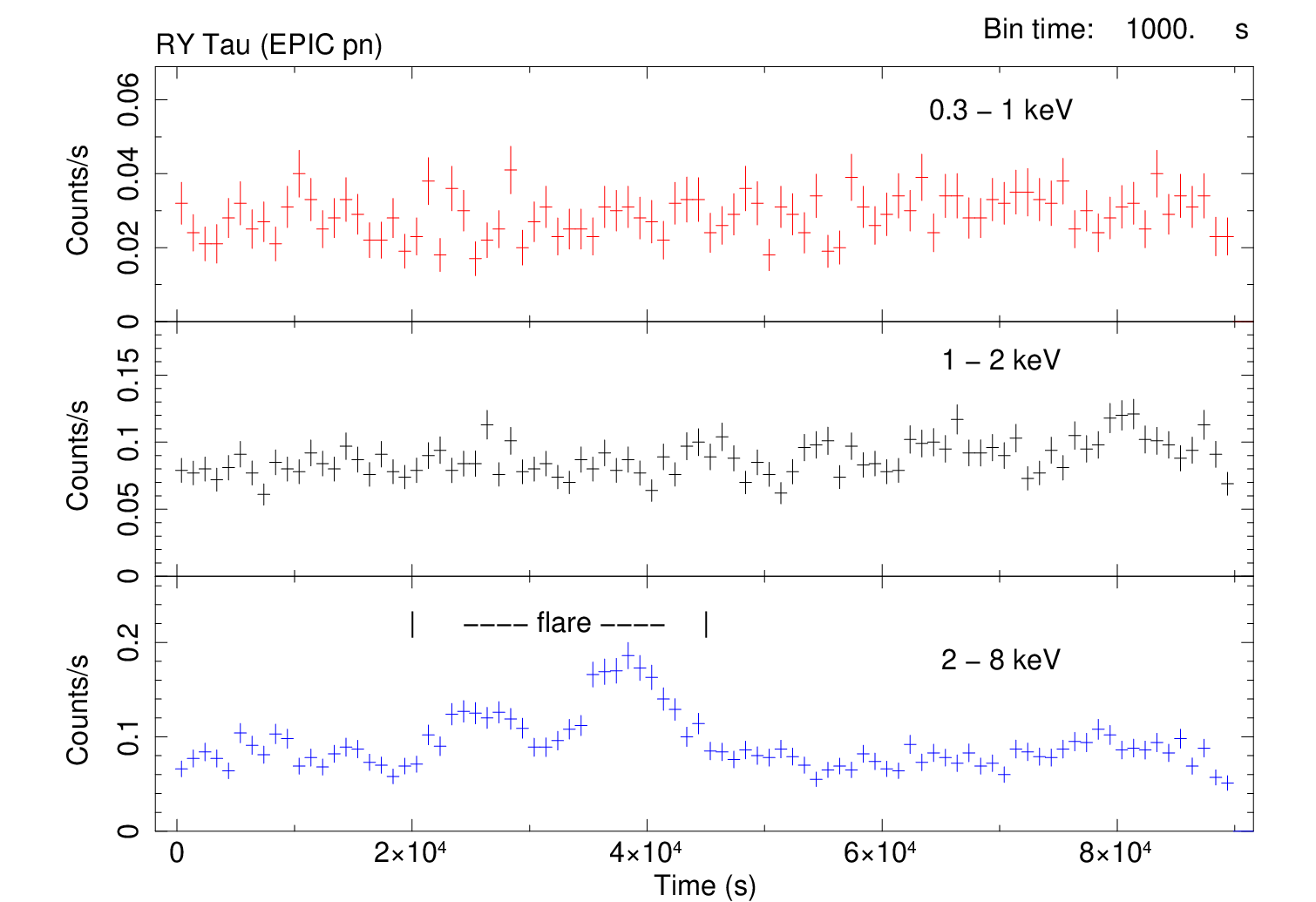}
\caption{Background-subtracted  XMM-Newton EPIC light curves of RY Tau (1000 s time bins).
Times are relative to the start of the observation.
~{\em Top}:~EPIC pn and MOS2 broad-band light curves (0.3 - 8 keV).
~{\em Bottom}:~EPIC pn light curves in three different energy bands.
No significant variability is present in the very soft energy band (0.3 - 1 keV) but the medium (1 - 2 keV)
and hard  (2 - 8 keV) band light curves are variable with high probability P$_{var}$ $>$ 0.99.
A separate spectrum was extracted for the interval marked `flare' in the hard-band  light curve.
}
\end{figure}
\clearpage

\begin{figure}
\figurenum{2}
\epsscale{1.0}
\includegraphics*[width=5.92cm,height=8.0cm,angle=-90]{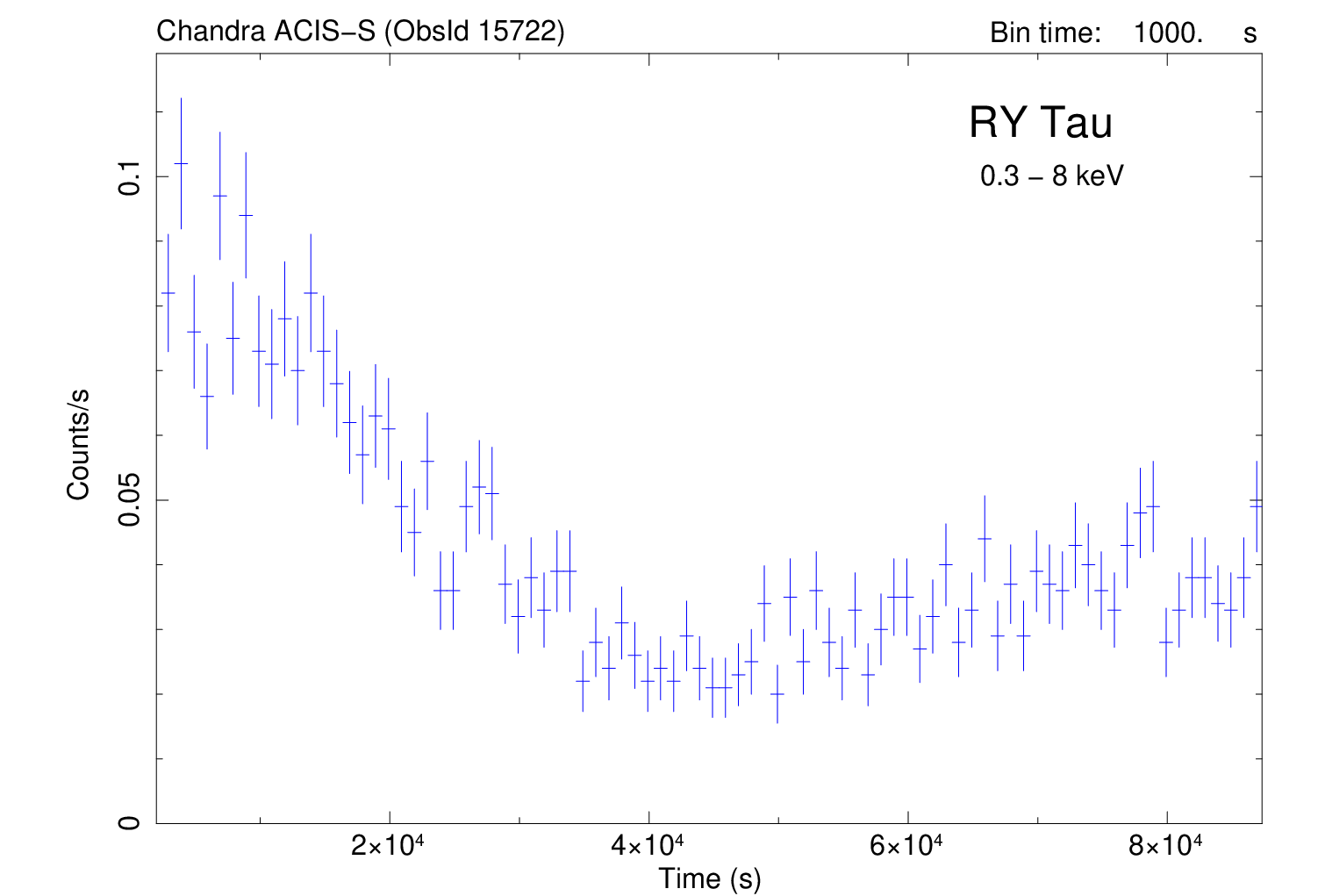}
\includegraphics*[width=5.92cm,height=8.0cm,angle=-90]{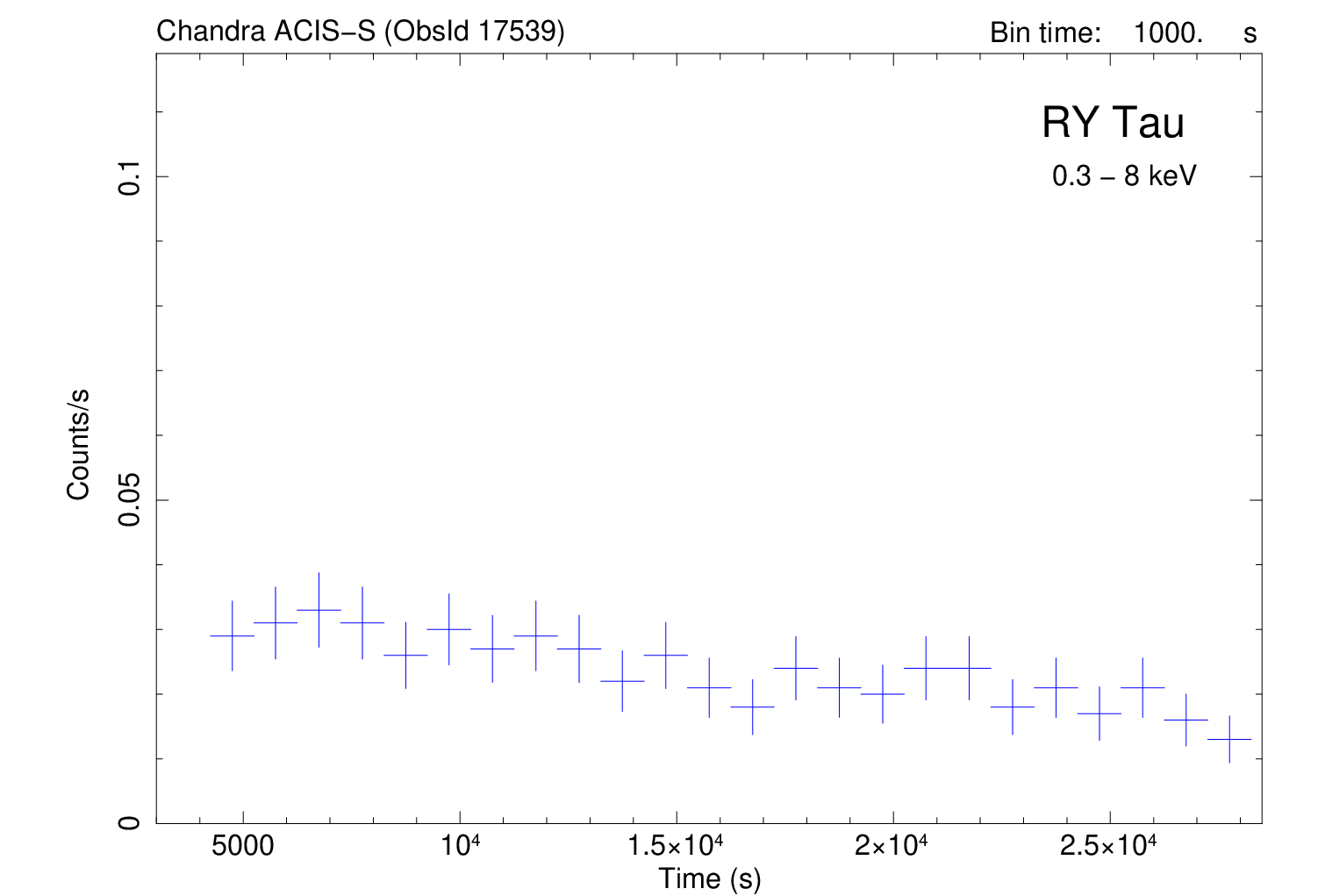} \\
\includegraphics*[width=5.92cm,height=8.0cm,angle=-90]{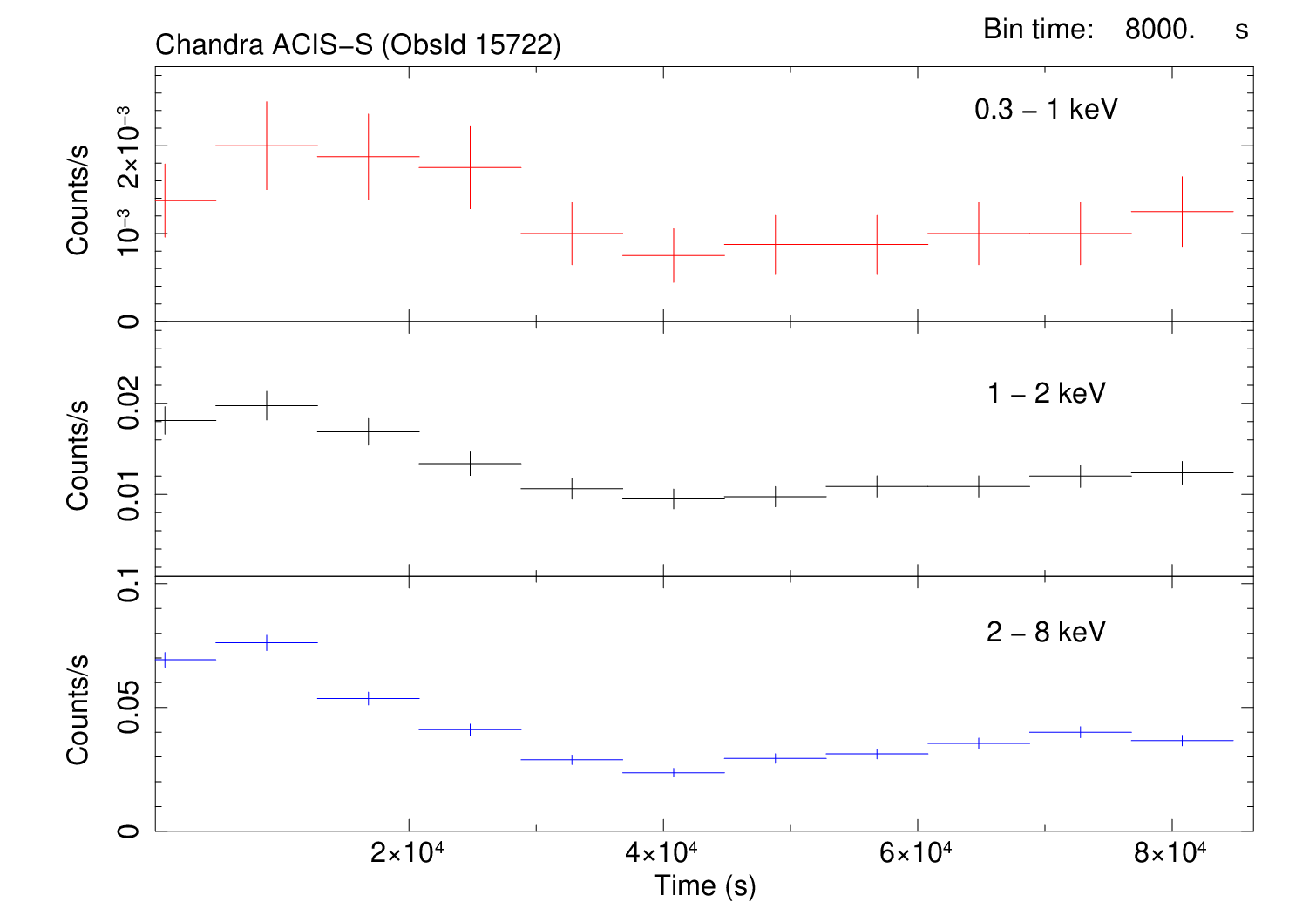} 
\includegraphics*[width=5.92cm,height=8.0cm,angle=-90]{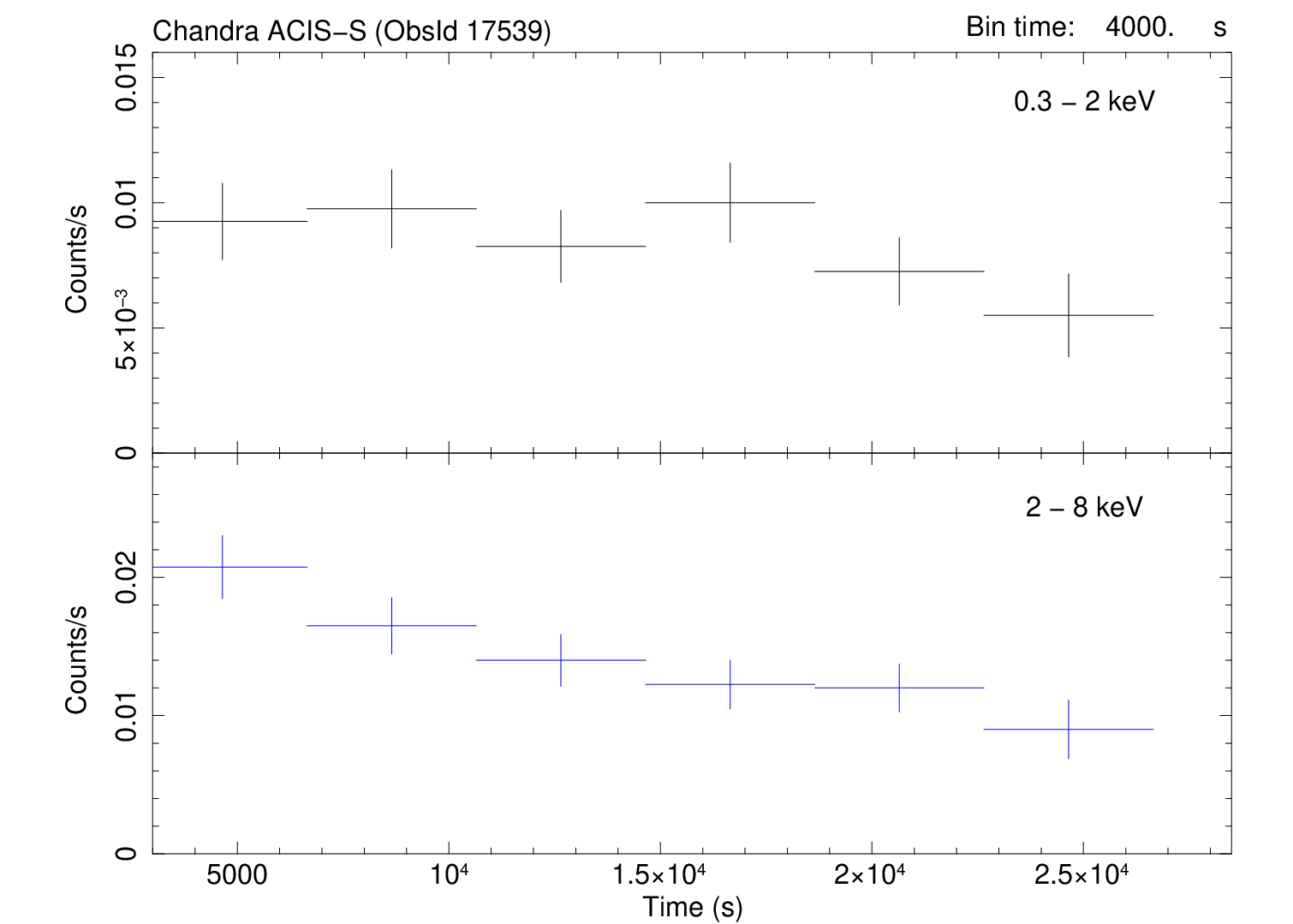} 
\caption{{\em Top}:~Chandra ACIS-S 0-order light curves (0.3 - 8 keV; 1000 s time bins) of RY Tau for
 ObsId 15722 (left) and 17539 (right). Times are relative to the observation start time.
 A gap of $\approx$18.5 hours occurred between the
 end of the first and beginning of the second observations.
~{\em Bottom Left}:~Chandra ACIS-S 0-order light curves of RY Tau in three different energy bands
for ObsId 15722 (8000 s time bins). Similar variability is present in very soft (0.3 - 1 keV),
medium (1 - 2 keV), and hard (2 - 8 keV) energy bands.
~{\em Bottom Right}:~Chandra ACIS-S 0-order light curves of RY Tau in the 0.3 - 2 keV and 2 - 8 keV  energy bands
for ObsId 17539 (4000 s time bins). There are insufficient counts to generate a very soft band (0.3 - 1 keV)
light curve.
}
\end{figure}
\clearpage

\begin{figure}
\figurenum{3}
\epsscale{1.0}
\includegraphics*[width=5.92cm,height=8.0cm,angle=-90]{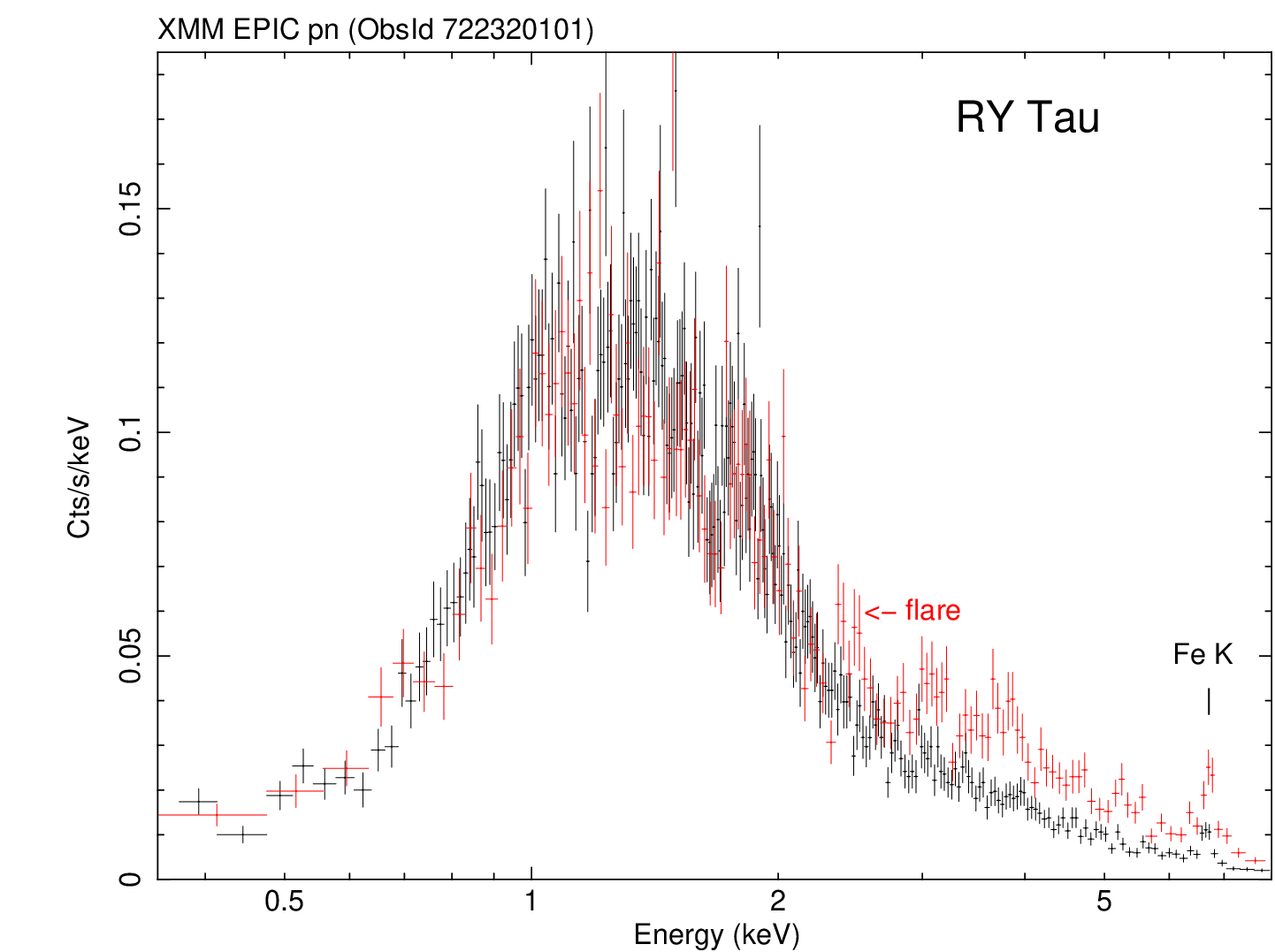} \\
\includegraphics*[width=5.92cm,height=8.0cm,angle=-90]{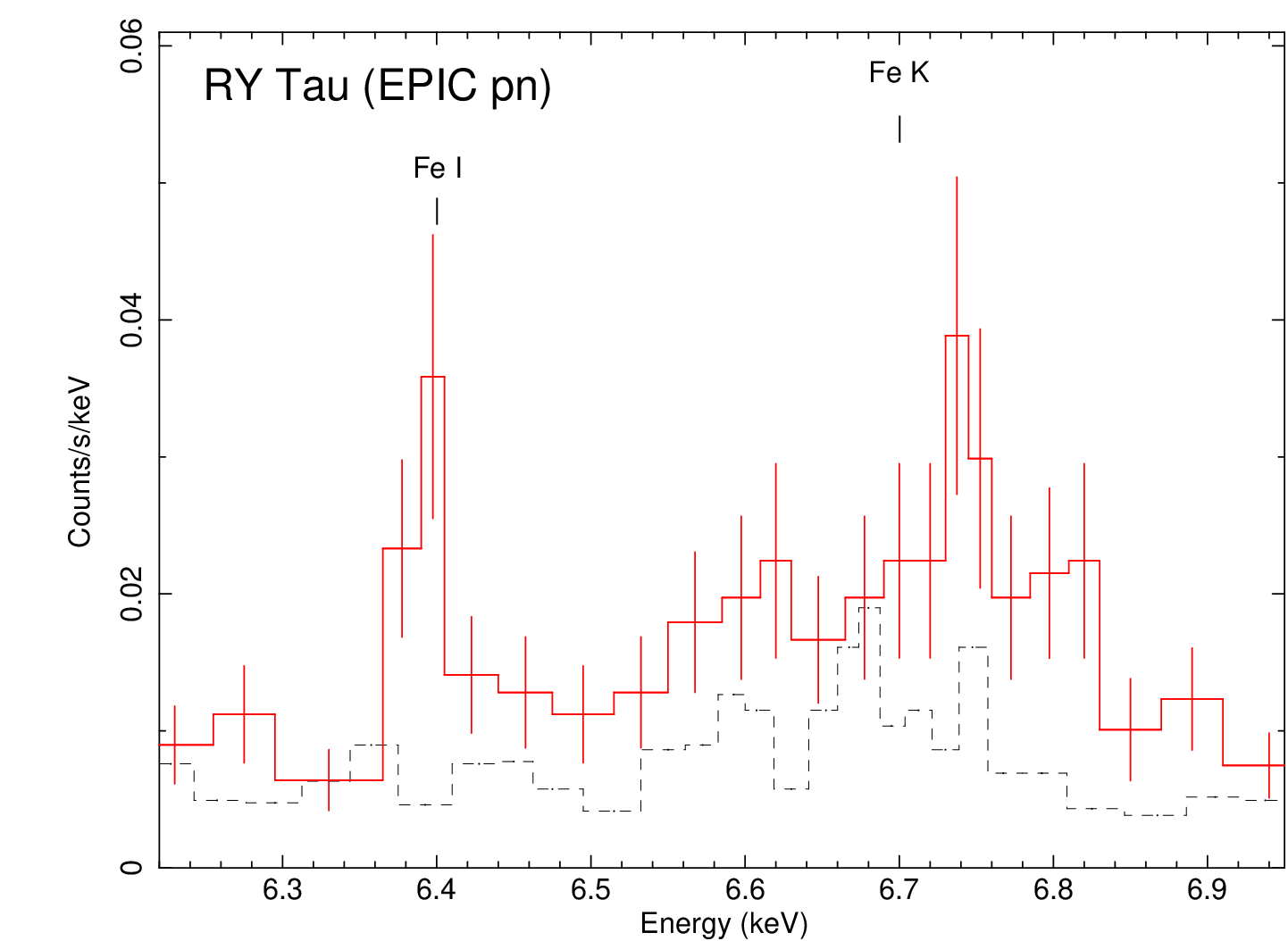}
\caption{{\em Top:}~XMM EPIC pn spectra of RY Tau binned to a minimum of 40 counts per bin.
The flare spectrum (red) includes only events recorded in the 25 ks flare segment (see Fig. 1).
The quiescent spectrum is based on
65 ks of exposure that excludes the flare. Both spectra exclude high background
data at the end of the observation.~
{\em Bottom:}~XMM EPIC pn flare (solid) and flare-excluded (dashed) spectra of RY Tau binned to a
minimum of 10 counts per bin.  The flare spectrum  shows emission from the Fe K line complex
which includes multiple Fe XXV transitions near 6.7 keV  arising in very hot plasma and
fluorescent Fe emission near 6.4 keV originating in cold irradiated material.
The EPIC pn energy resolution at 6.7 keV is FWHM $\approx$ 0.15 keV.
}
\end{figure}
\clearpage

\begin{figure}
\figurenum{4}
\epsscale{1.0}
\includegraphics*[width=5.92cm,height=8.0cm,angle=-90]{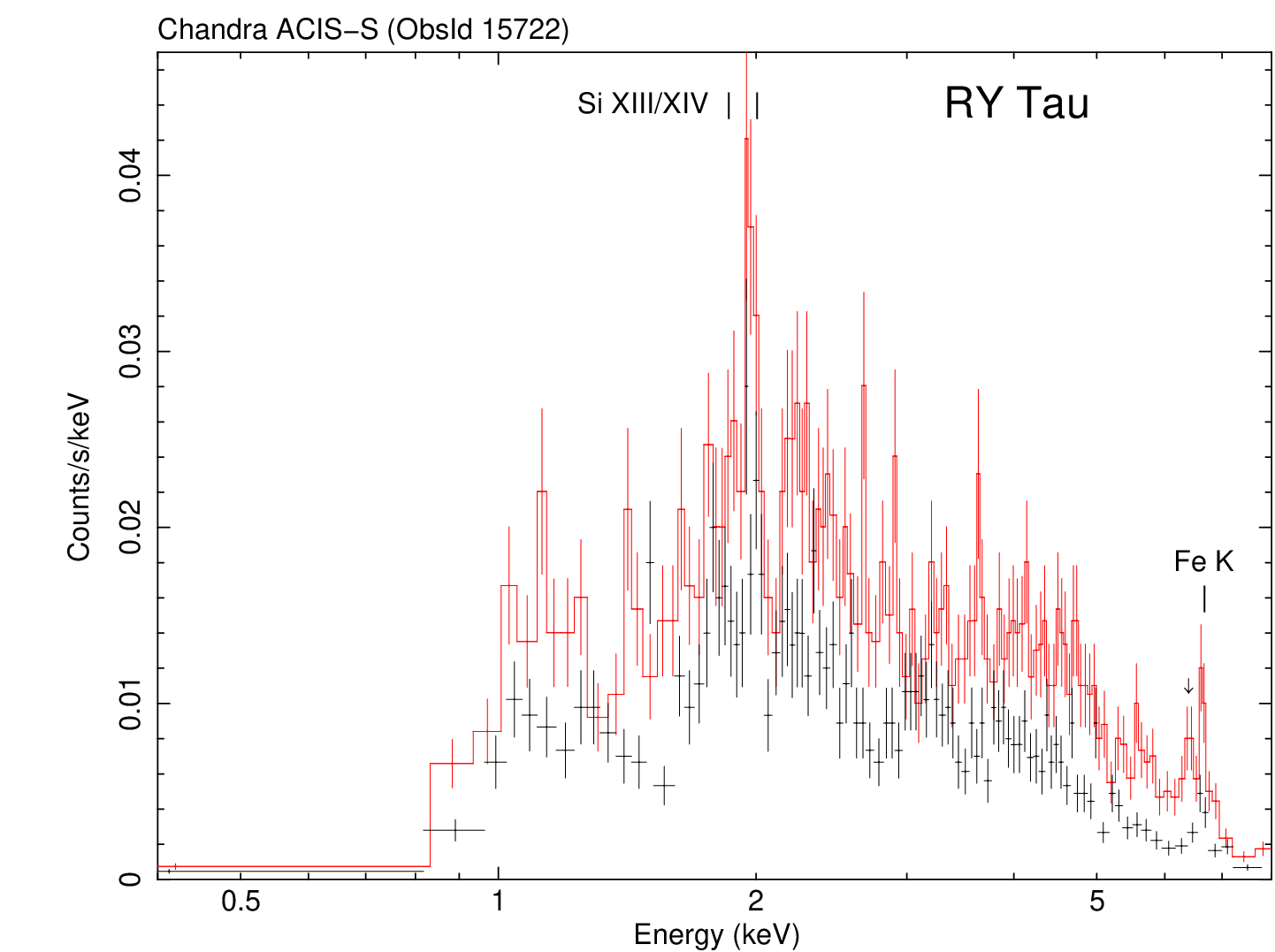} \\
\includegraphics*[width=5.92cm,height=8.0cm,angle=-90]{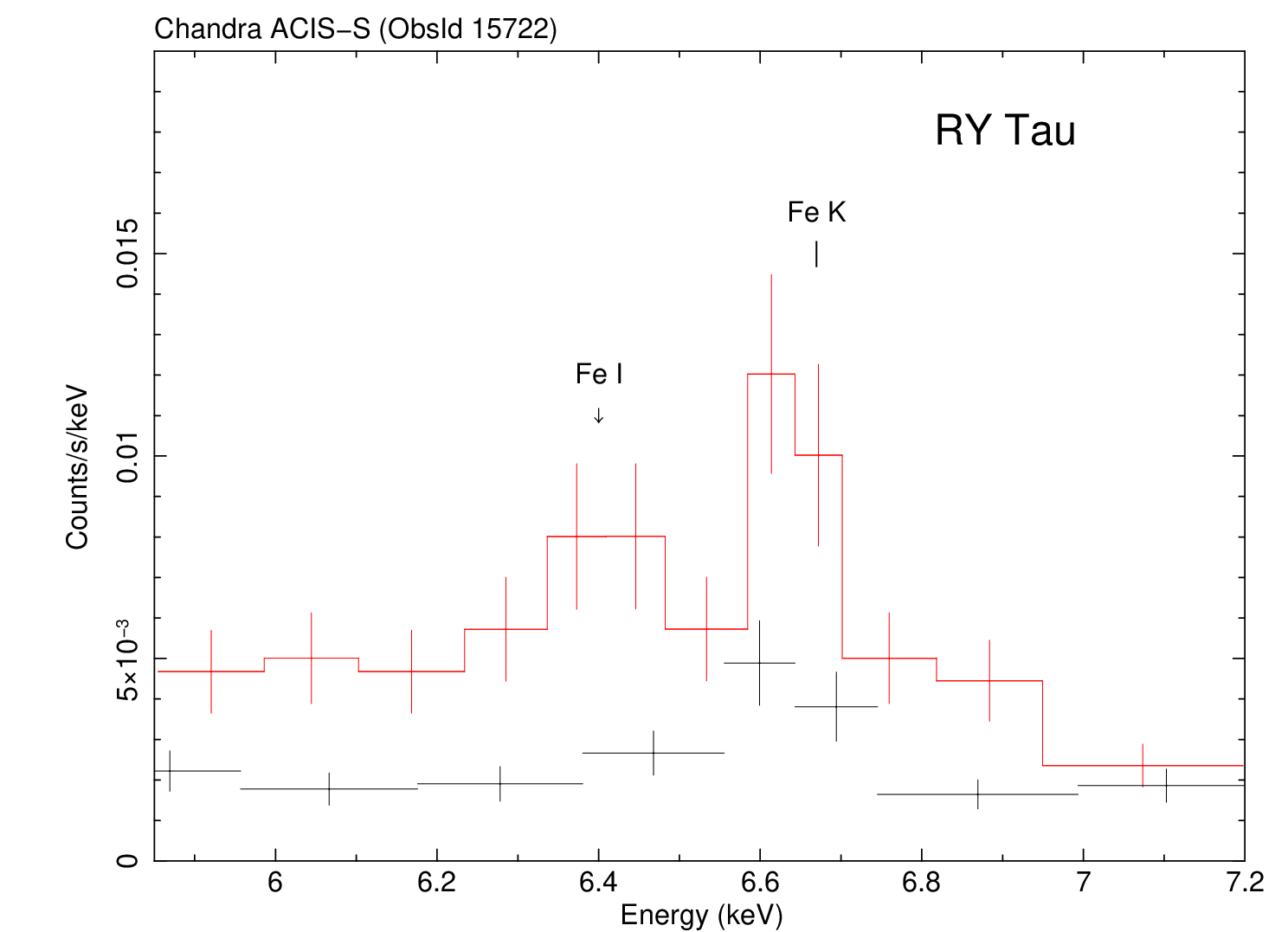}
\caption{{\em Top}:~Chandra ACIS-S 0-order spectra of RY Tau for
 ObsId 15722 binned to a minimum of 20 counts per bin. The spectrum connected by the
 solid histogram (red) consists of  2763 counts recorded in the first 35 ks of the 
 observation (high-state)
 when the count rate was high and slowly decaying.
 The spectrum without the histogram line (black) consists of 2338 counts recorded
 during the low-state from t = 35 ks to the end of the observation (duration 52.586 ks). 
 Arrow marks probable fluorescent
 Fe emission at 6.4 keV.
 {\em Bottom}:~Same as top  showing a more detailed view of the Fe lines  between 6.4 - 6.7 keV.
               The ACIS chip S3 energy resolution from pre-launch measurements is
               FWHM $\approx$ 0.2 keV at 6.7 keV.}
\end{figure}
\clearpage

\begin{figure}
\figurenum{5}
\epsscale{1.0}
\includegraphics*[width=5.92cm,height=8.0cm,angle=-90]{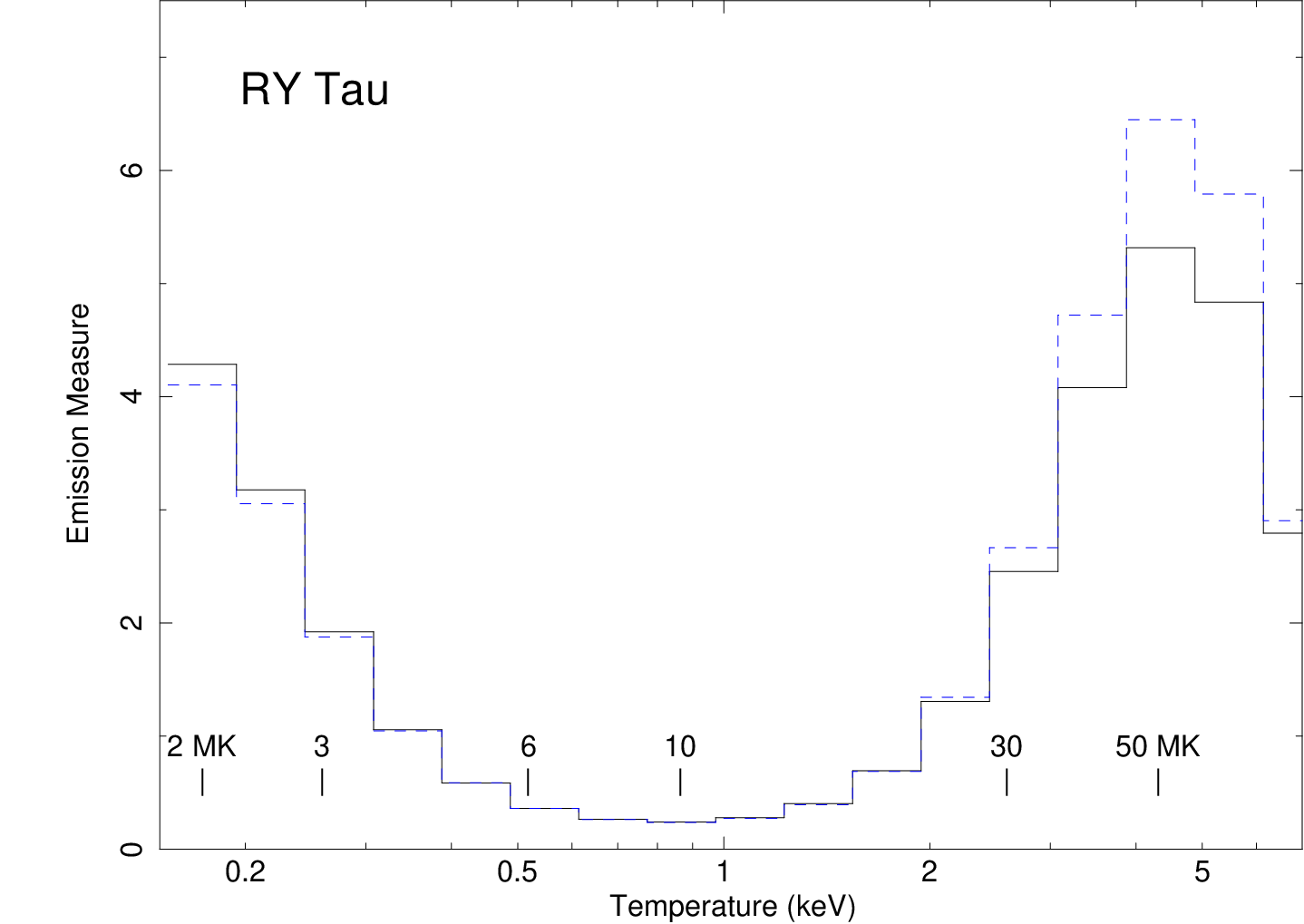} \\
\includegraphics*[width=5.92cm,height=8.0cm,angle=-90]{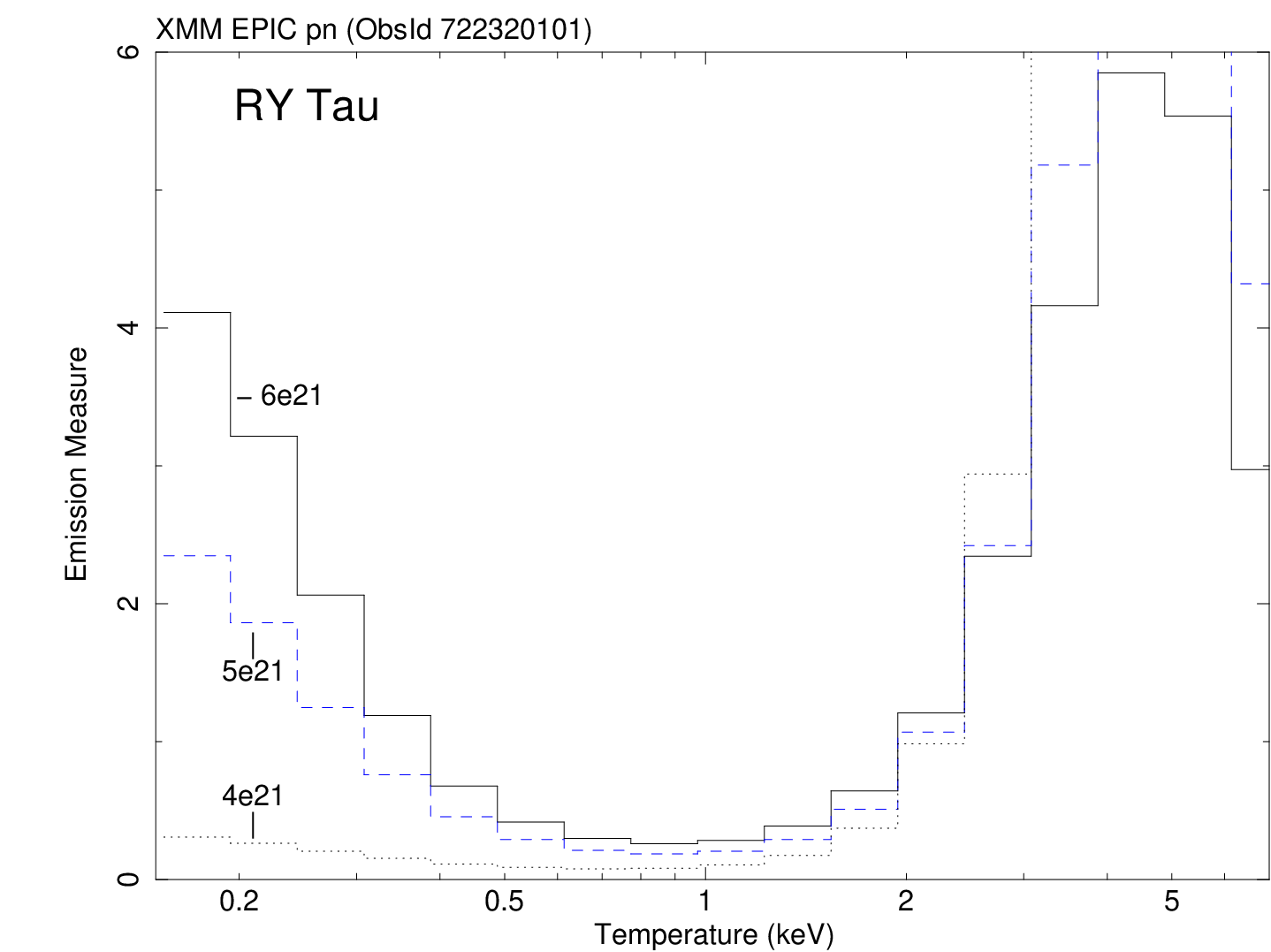}
\caption{{\em Top:}~Emission measure distribution of RY Tau obtained by fitting {\em XMM-Newton} spectra
excluding the  flare intervals.  The fits were obtained with the  XSPEC variable abundance
model $c6pvmekl$ model (Table 4).
The solid histogram is based on a fit of the EPIC pn spectrum and dashed histogram
is based on a simultaneous fit of EPIC pn $+$ RGS1\&2 spectra. The best-fit
absorption is N$_{\rm H}$ = 6.0 $\times$ 10$^{21}$ cm$^{-2}$.
{\em Bottom:}~A comparison of the emission measure distributions obtained by fitting
the EPIC pn spectrum with the $c6pvmekl$ model using three different fixed
values of N$_{\rm H}$. Flare intervals were excluded. All three fits are acceptable
with respective $\chi^2_{\nu}$ values 0.98 (N$_{\rm H}$ = 6e21 cm$^{-2}$),
0.99 (N$_{\rm H}$ = 5e21 cm$^{-2}$), and 1.02 (N$_{\rm H}$ = 4e21 cm$^{-2}$).
}
\end{figure}
\clearpage

\begin{figure}
\figurenum{6}
\epsscale{1.0}
\includegraphics*[width=7.40cm,height=10.0cm,angle=-90]{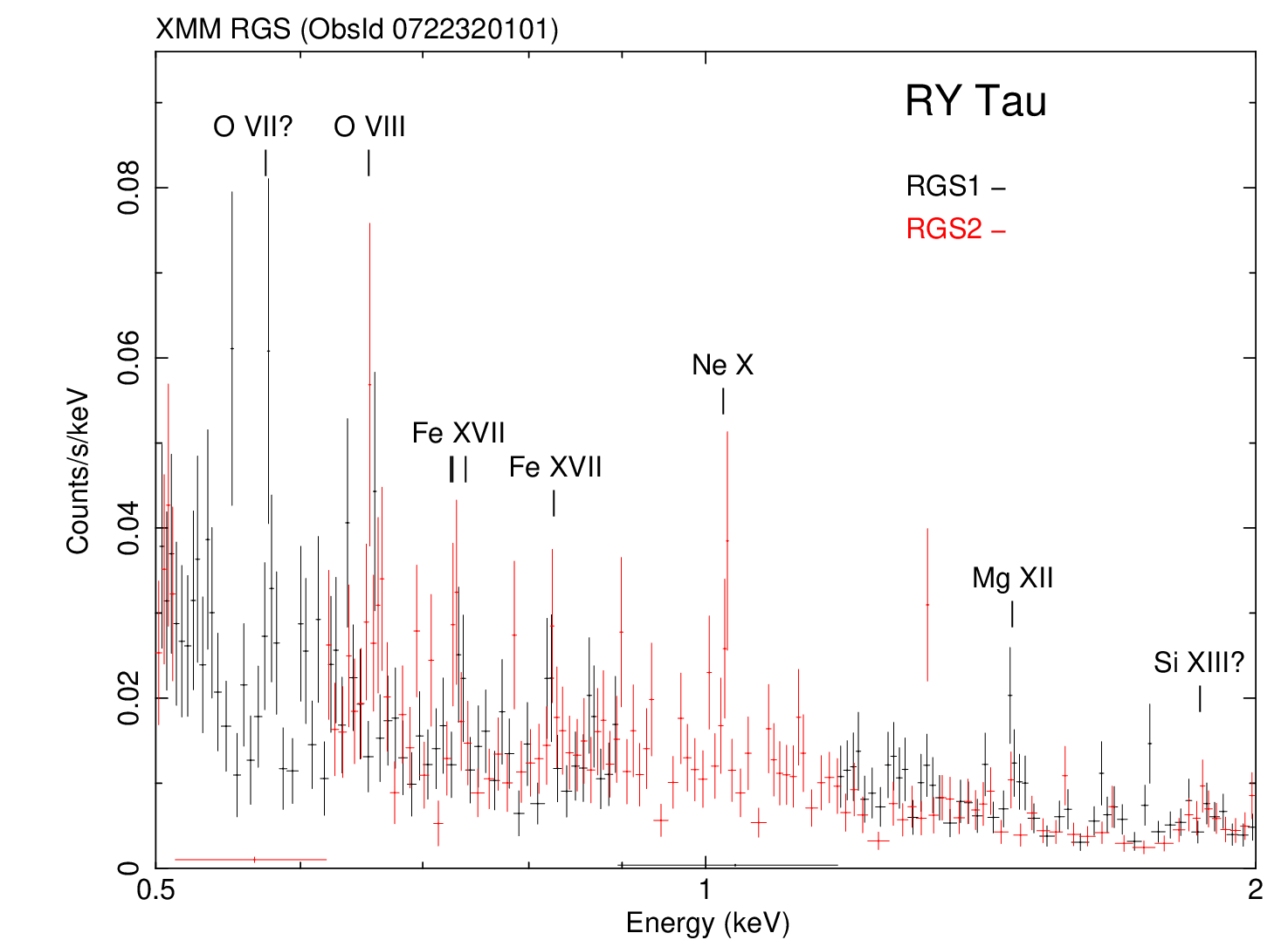} \\
\includegraphics*[width=7.40cm,height=10.0cm,angle=-90]{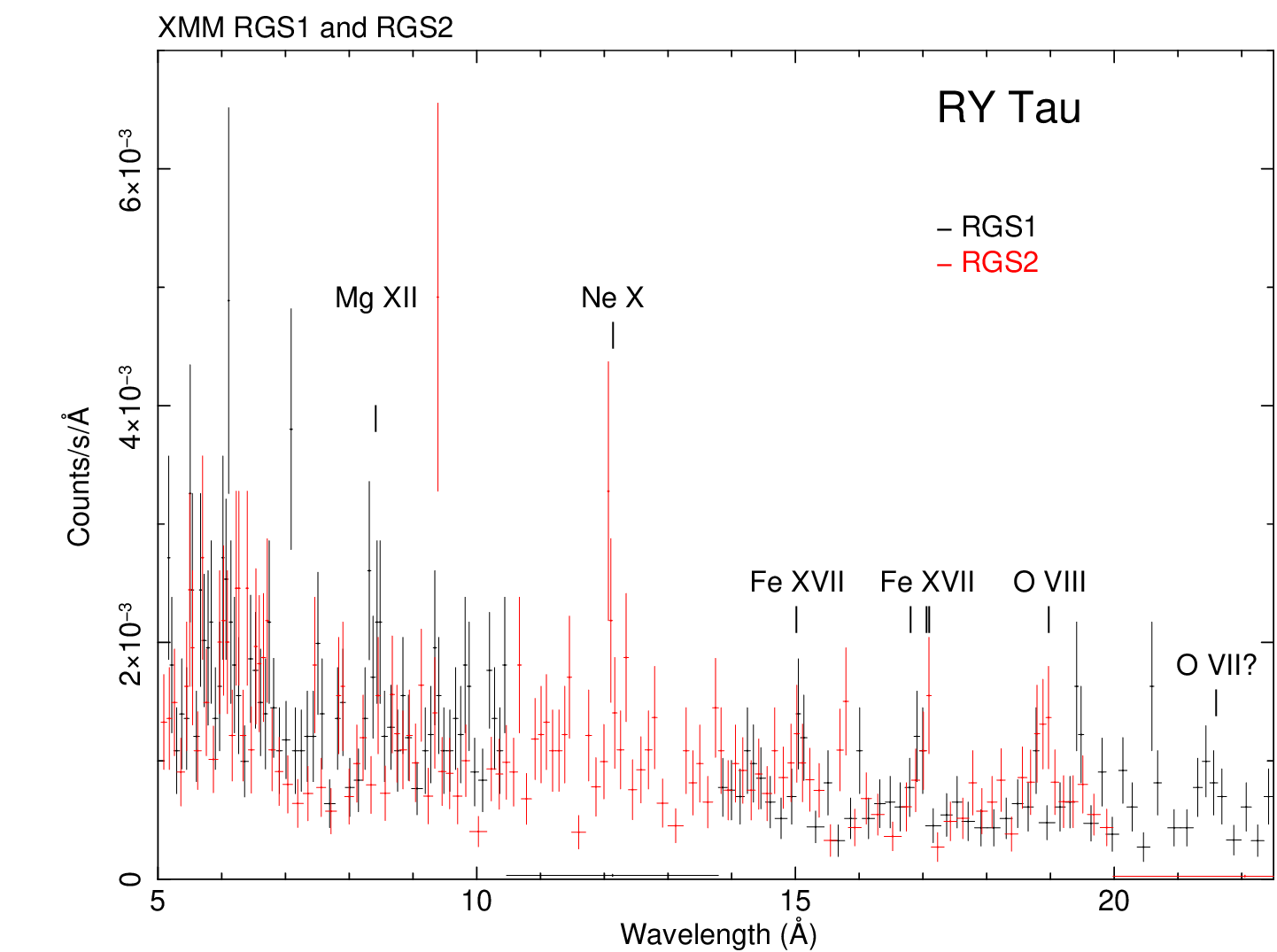} \\
\caption{~{\em Top}:~Binned XMM-Newton 1st order RGS1 (black) and RGS2 (red) spectra of RY Tau
          based on the first 92.25 ks of exposure, plotted versus energy. Due to the faint emission, no
          background subtraction has been applied. Background dominates below $\approx$0.5 keV.
         Coverage gaps  due to CCD failures are
         present at 0.9 - 1.17 keV (RGS1) and 0.51 - 0.62 keV (RGS2).
         Vertical lines mark  line reference energies.
~{\em Bottom}:~Same as above only plotted versus wavelength. The identification of the RGS1 
         feature at 21.488 \AA ~(E = 0.577 keV) as O VII$r$ is questionable (see text). 
         }
\end{figure}
\clearpage

\begin{figure}
\figurenum{7}
\epsscale{1.0}
\includegraphics*[width=5.92cm,height=8.0cm,angle=-90]{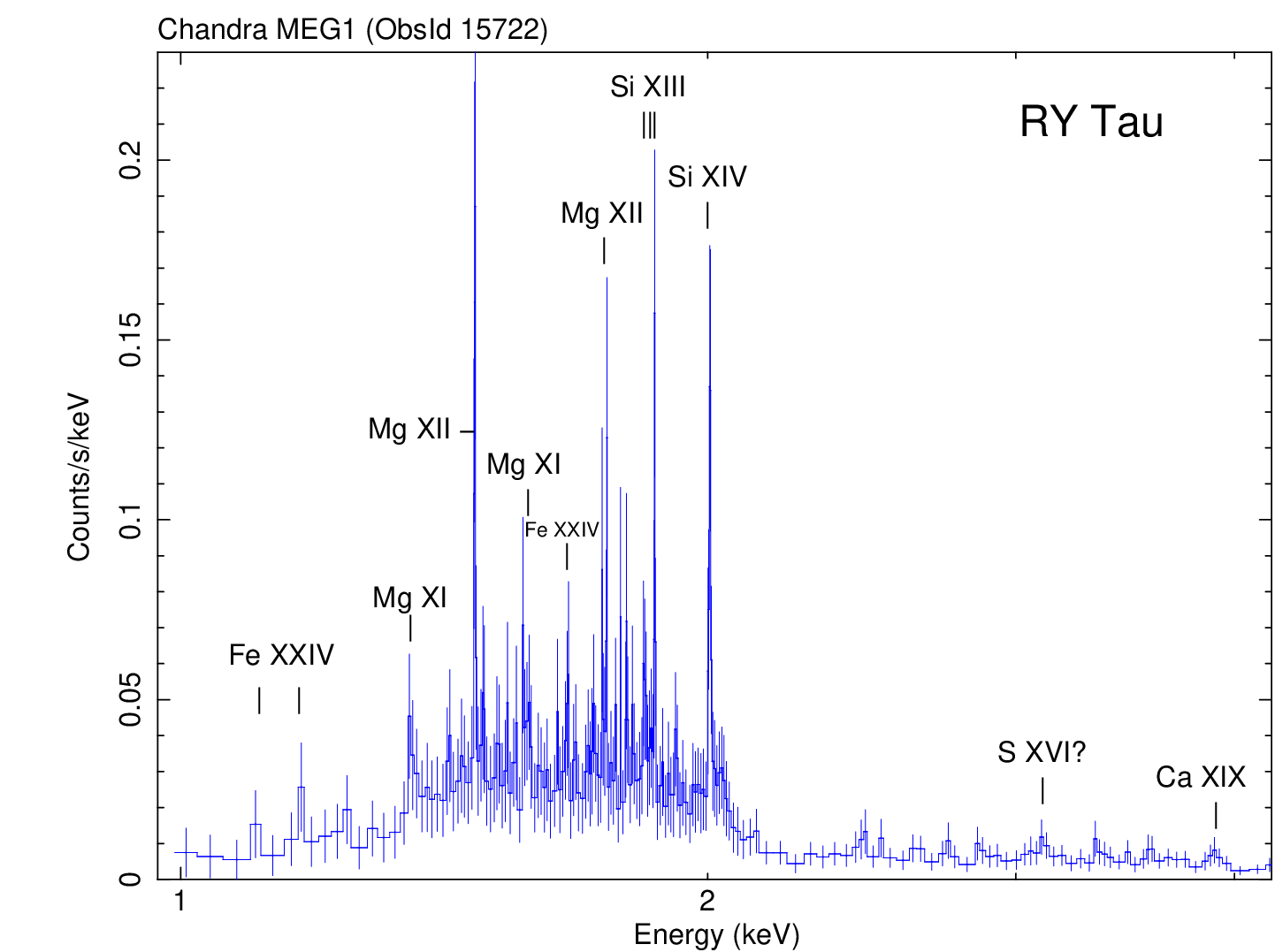}
\includegraphics*[width=5.92cm,height=8.0cm,angle=-90]{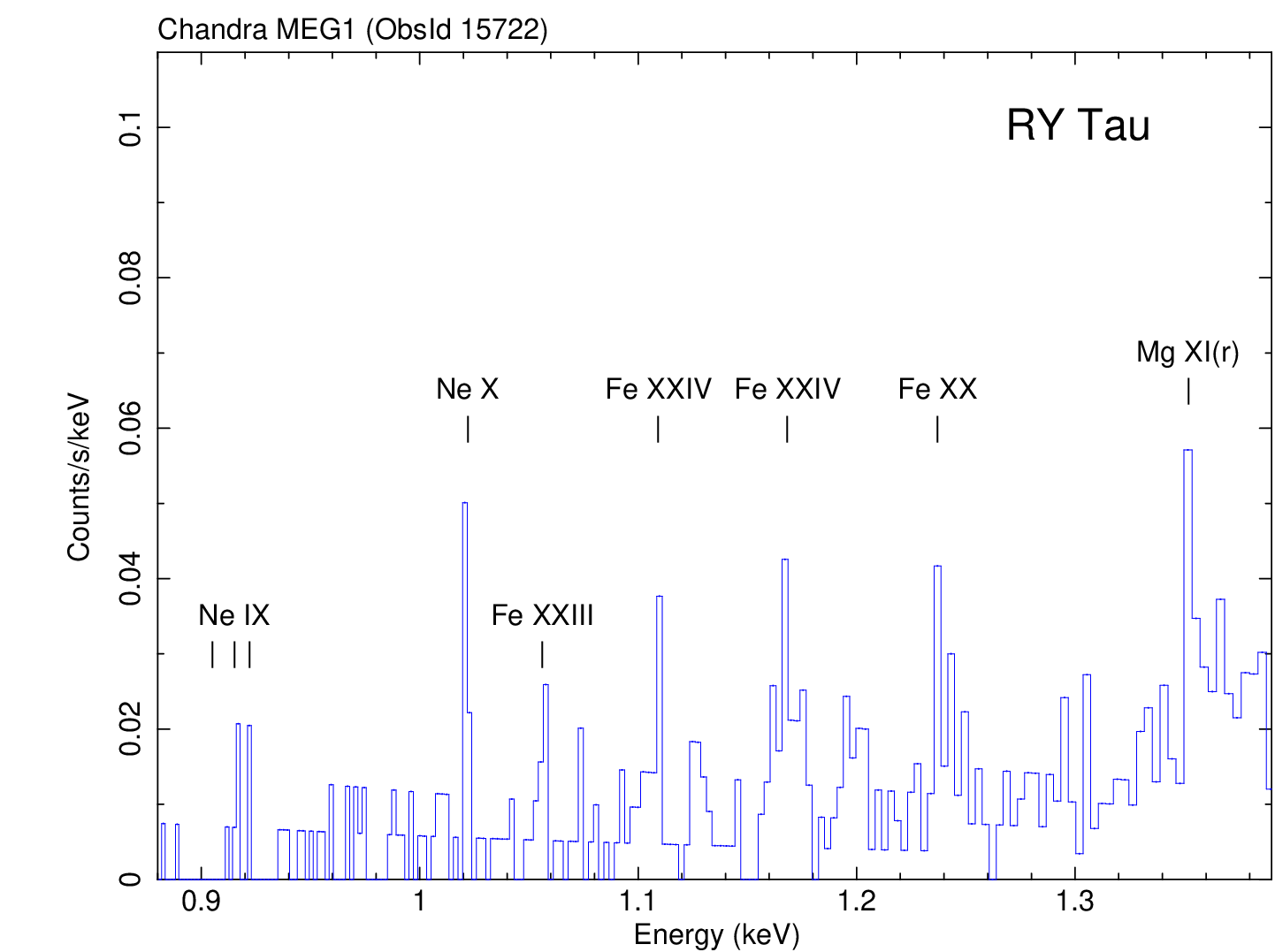} \\
\includegraphics*[width=5.92cm,height=8.0cm,angle=-90]{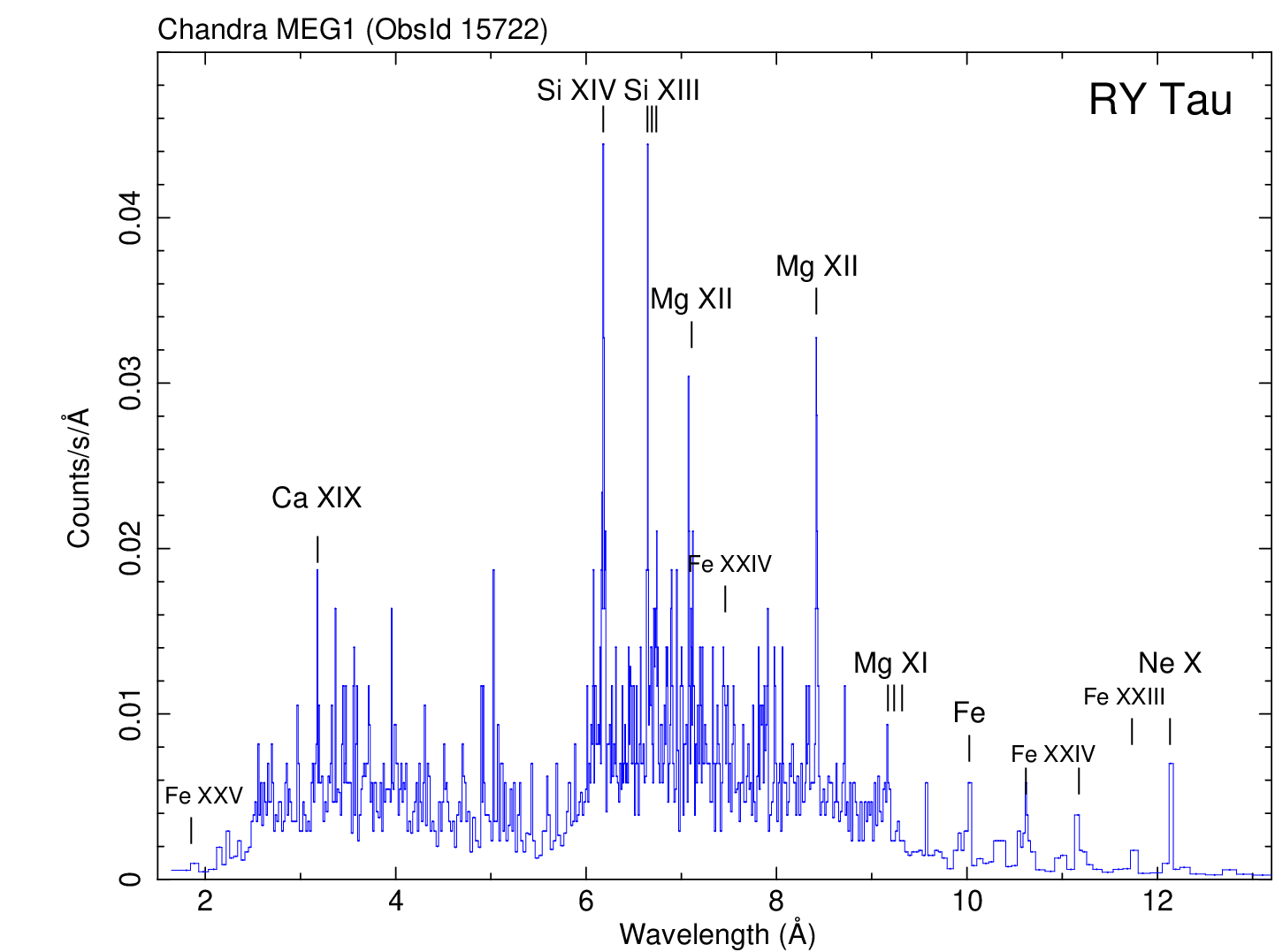}
\includegraphics*[width=5.92cm,height=8.0cm,angle=-90]{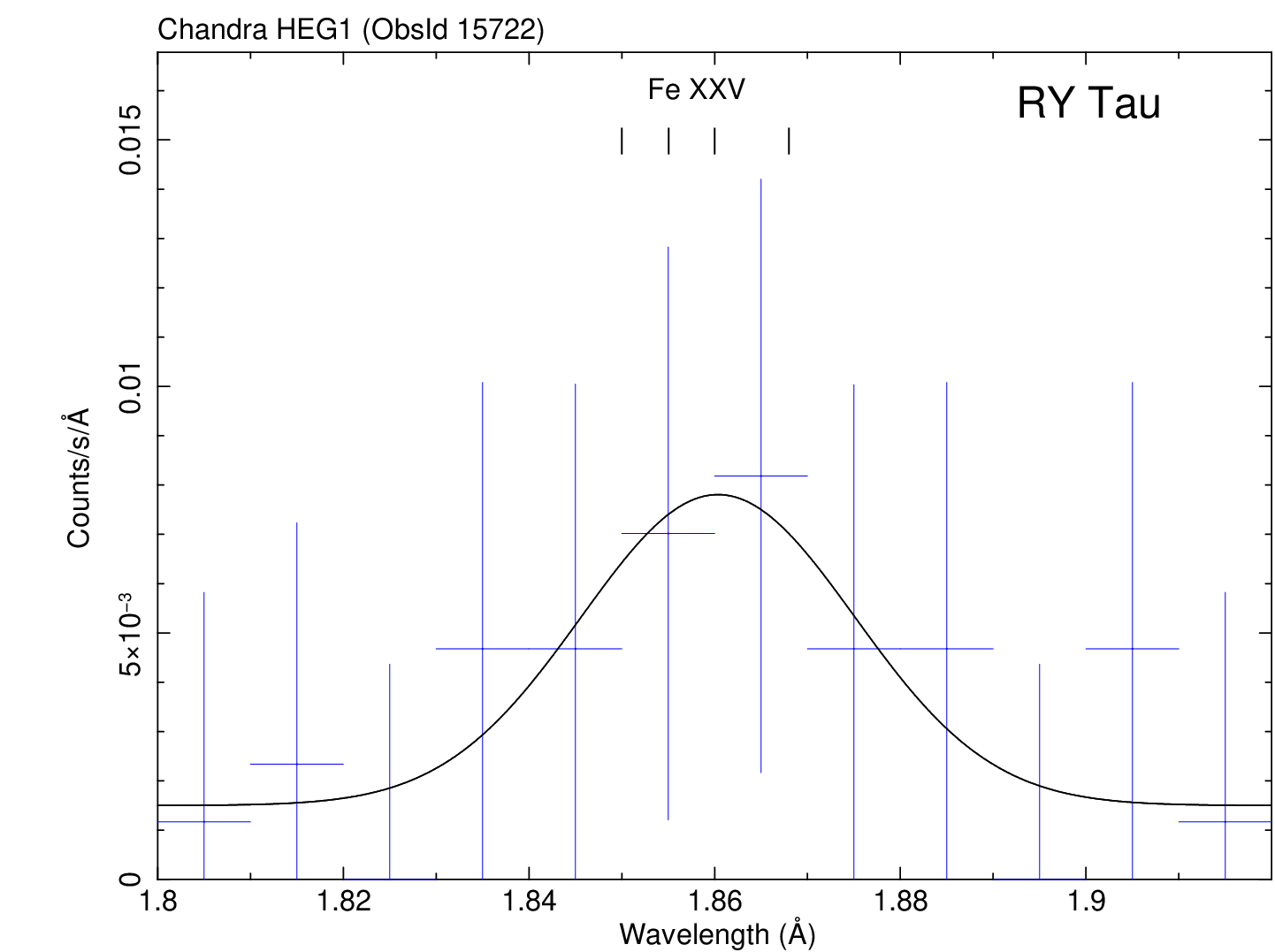} \\
\caption{Lightly-binned Chandra 1st-order grating spectra ($+$1 and $-$1 orders combined) of
RY Tau for the full ObsId 15722 exposure (85.5 ks). Vertical lines mark reference
energies/wavelengths.
{\em Top}:~MEG1 plotted versus energy. Error bars omitted for clarity on low-energy
portion at right.
{\em Bottom Left}:~MEG1 plotted versus wavelength. Error bars omitted for clarity.
~{\em Bottom Right}:~HEG1 showing the  Fe K$\alpha$ line complex with
reference wavelengths of four Fe XXV transitions  marked.
The fitted Gaussian line width (FWHM = 0.034 \AA) is
  broader than the instrumental width and indicates contributions from multiple Fe XXV lines.
}
\end{figure}
\clearpage

\begin{figure}
\figurenum{8}
\epsscale{1.0}
\includegraphics*[width=5.92cm,height=8.0cm,angle=-90]{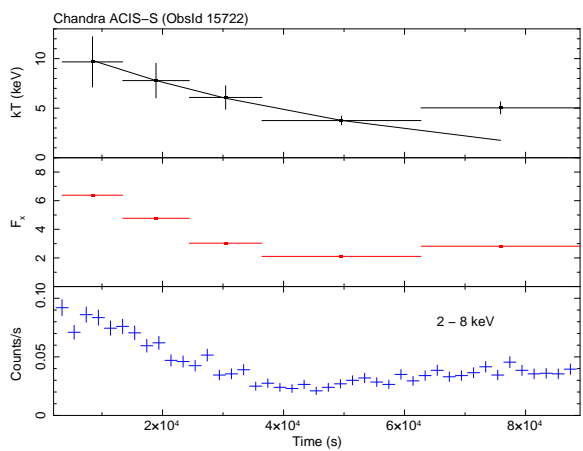}
\caption{Evolution of mean plasma temperature (kT), absorbed X-ray flux F$_{x}$ (0.2 - 10 keV;
 in units of 10$^{-12}$ ergs cm$^{-2}$ s$^{-1}$), and ACIS-S
 hard-band (2 - 8 keV)  count rate versus time for RY Tau during {\em Chandra} ObsId 15722.
 The values of kT and F$_{x}$ were determined by fitting time-partitioned ACIS-S spectra
 binned to a minimum of 10 counts per bin using a 1T $apec$ model with metallicity fixed at
 $Z$ = 0.4 $Z_{\odot}$. The horizontal error bars in the top two panels demarcate  the time intervals used to
 extract time-partitioned spectra. The solid line in the top panel is an exponential fit of the first
 four data points with an e-folding time of 52.59 ks.
}
\end{figure}
\clearpage

\end{document}